\pdfoutput=1

\documentclass[conference,compsoc]{IEEEtran}

\newif\ifanon
\anonfalse

\usepackage[T1]{fontenc}
\usepackage[utf8]{inputenc}
\usepackage{graphicx}
\usepackage{amsmath}
\usepackage{amssymb}
\usepackage{booktabs}
\usepackage{multirow}
\usepackage{xcolor}
\usepackage[caption=false,font=footnotesize]{subfig}
\usepackage{algorithm}
\usepackage[noend]{algpseudocode}
\floatevery{algorithm}{\scriptsize}
\makeatletter\renewcommand\floatc@ruled[2]{\scriptsize{\@fs@cfont #1} #2\par}\makeatother
\algrenewcommand\alglinenumber[1]{\scriptsize #1:}
\usepackage{xspace}
\usepackage{cite}
\usepackage[hidelinks]{hyperref}
\usepackage{cleveref}

\newcommand{\sysname}{{\texttt{obfsgen}}\xspace}
\newcommand{\eg}{e.g.,\xspace}
\newcommand{\ie}{i.e.,\xspace}

\begin{document}
\bstctlcite{BSTcontrol}

\title{\sysname: Generative Tunnel Obfuscation}

\ifanon
  \author{\IEEEauthorblockN{Anonymous Submission}}
\else
  \author{\IEEEauthorblockN{Kaidong Wu\textsuperscript{1}, Wentao Du\textsuperscript{1}, Shinan Liu\textsuperscript{2}, Diwen Xue\textsuperscript{1}}
    \IEEEauthorblockA{\textsuperscript{1}The Chinese University of Hong Kong}
    \IEEEauthorblockA{\textsuperscript{2}The University of Hong Kong}}
\fi

\maketitle

\begin{abstract}
Encrypted tunnels are widely used to circumvent censorship, which has drawn censors' efforts to detect and block them.
As the circumvention tunnels in widest use have largely converged on TLS, detection has shifted from the content of their traffic to its \textit{shape}, where recent research and real-world blocking events have revealed exploitable fingerprints of tunneling.
Existing traffic-shaping obfuscators such as \texttt{obfs4}, however, are fundamentally limited in hiding these fingerprints, as their \textit{application-driven} design can only add to the shape that the application creates.
A \textit{schedule-driven} obfuscator, which decides when and how much to send independently of the application, can give tunnels any shape but raises new challenges, first and foremost \textit{what a good shape is}.

This paper presents \sysname, a schedule-driven obfuscator for tunneling traffic.
\sysname samples its schedule from a generative sequence model of \textit{normal} traffic, which predicts the future shape of a flow from its whole history, to capture the long-range dependencies that make a flow's shape normal.
Built around this model, \sysname addresses system problems that such an obfuscator faces in practice, such as synchronization across a noisy network.
We evaluate the (in)distinguishability that \sysname offers, the performance overhead that it adds and the computational resources that it requires.
We see the schedule-driven architecture as the next generation of tunnel obfuscators, and we hope that \sysname and the rationale behind its design help make such obfuscators practical and deployable.
\end{abstract}

\section{Introduction}
\label{sec:intro}

\suppressfloats[t]
\begin{figure}[t]
\centering
\resizebox{\columnwidth}{!}{\includegraphics{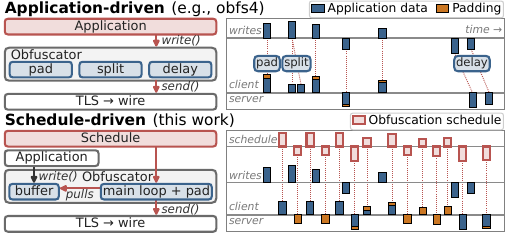}}
\caption{Application-driven vs. schedule-driven obfuscation.}
\label{fig:driven}
\end{figure}

Encrypted tunnels, such as VPNs and encrypted proxies, are important tools for online privacy, and increasingly also for circumventing censorship to reach information that a user's network blocks.
A tunnel protects its user by limiting what parties along the network path can see, be they nation-state censors~\cite{khattak2016sok,raman2020censoredplanet,niaki2020iclab}, surveillance agencies~\cite{abelson2015doormats,adrian2015imperfect}, or commercial ISPs that monetize their customers' traffic data~\cite{ftc2021isp}.
This loss of visibility gives these parties an incentive to detect, discriminate against or outright restrict tunneling traffic.
Nation-state censors around the world have been blocking circumvention tunnels for more than a decade~\cite{winter2012tor,ensafi2015probing,wu2023gfw,xue2022tspu,ooni2022iran}, and commercial ISPs have asked regulators to restrict their use, citing the loss of traffic data~\cite{mobileuk2022cma,eutelecomceo2022letter}.
In such networks, obfuscating a tunnel (\ie concealing the characteristics of tunneling) is therefore no longer an optional privacy feature but close to a precondition for the tunnel to work \textit{at all}.

After a decade-long arms race between censors and obfuscators~\cite{tschantz2016sok,xuebridging2024}, the circumvention tunnels in widest use today (\eg Trojan, VLESS, NaiveProxy~\cite{trojangithub,xraygithub,naiveproxygithub}) have largely converged on TLS, whose ubiquity on the Internet gives them strong cover.
Because these tunnels borrow from mainstream TLS implementations---those of popular browsers~\cite{frolov2019tlscensorship}---a censor that inspects the bytes of their packets finds little that sets them apart from ``normal'' TLS such as HTTPS.
The arms race has therefore shifted from the content of the traffic to its \textit{shape}, \ie the sizes, timing and directions of its packets.
In 2022, China's Great Firewall (GFW) started to block TLS-based tunnels~\cite{gfwreport2022tlsblocking}, and academic research has since identified several fingerprints in the shape of their traffic that tell such tunnels apart from normal TLS~\cite{xue2024encapsulated,xue2025timing,ramesh2024calculatency,ali2026beyondrtt}.
In response, circumvention tunnels have, over the past three years, released various patches that target these fingerprints as each is discovered~\cite{xraygithub,anytlsgithub,restlsgithub}, most often by adding arbitrary padding (and sometimes delay) in an attempt to obscure the shape of the traffic.

We argue that, by design, most such patches cannot fully hide the fingerprints that tunneling creates.
Existing obfuscators and their patches follow what we call an \textit{application-driven} design (Fig.~\ref{fig:driven}, top), in which each \texttt{write()} of the application triggers the obfuscator to pad, split or delay the data written.
Such a design can \textit{add} padding and delay, but it is fundamentally unable to act when the application is idle, for example to send traffic \textit{before} the application's next \texttt{write()}, in order to hide the extra round-trip time induced by a tunnel~\cite{xue2025timing}.
A better architecture, we argue, is a \textit{schedule-driven} design (Fig.~\ref{fig:driven}, bottom), in which the obfuscator sets its own schedule of when to send and how much each record carries, independently of the application.
Such a design has the flexibility to give tunneling traffic any shape required for obfuscation.

Designing a schedule-driven obfuscator, however, raises several challenges of its own, first and foremost deciding \textbf{what a good shape is}.
Existing application-driven obfuscators defer this decision to the application, which largely determines the shape of the tunnel's traffic.
A schedule-driven obfuscator that aims to blend tunneling traffic into ``normal'' traffic, on the other hand, must first understand \textit{what normal means in practice}, and then determine how such normal traffic can be captured and reproduced faithfully on the wire.
Despite practitioners' calls for a better obfuscation architecture, these challenges have impeded the development of a practical schedule-driven obfuscator.

This paper describes our attempt to answer these questions, along with \sysname{}, a schedule-driven obfuscator for tunneling traffic.
We believe that what is \textit{normal} varies with the vantage point of the censor and with the network that it monitors, which encourages a modular approach in which each deployment chooses its own \textit{reference} for normal, \ie the traffic that its censor would compare the tunnel's traffic against.
As an example reference, we collected background traffic from a gateway of our campus network over a month (\S~\ref{sec:dataset}).
\sysname{} then captures the reference by \textit{learning} a generative sequence model from it.
The model predicts the shape of each future TLS record of a flow from all the records seen before it (\S~\ref{sec:model}).
Unlike prior approaches that use marginal or low-order Markov statistics~\cite{wright2009morphing,moghaddam2012skypemorph, censorkl}, the sequence model conditions on the whole history to capture the long-range dependencies that make each flow's shape ``normal''.
\sysname{} is then built around this generative model, solving the challenges that an obfuscator faces in the real world, such as synchronization (\S~\ref{sec:system-history}), deadlocks and race conditions (\S~\ref{sec:system-async}), so that two \sysname{} endpoints can realize an ideal shape together, in real time, across a lossy and noisy Internet.

We evaluate \sysname{} under a variety of network conditions, for its (in)distinguishability as well as its practicality.
We first establish that the negative log-likelihood (NLL) of a flow under our model tracks distinguishability across various deviations from normal well enough to serve as a proxy for it (\S~\ref{sec:model-proxy}), and then evaluate distinguishability directly, with classifiers of different types under varying threat models.
\sysname{} reduces a \textit{blocklisting} censor, which looks for known tunneling traffic, to little better than random guessing.
Against an \textit{allowlisting} censor, which flags anything that deviates from normal, \sysname{} also provides reasonable obfuscation, without knowing in advance what the censor looks for.
However, \sysname{} traffic remains moderately distinguishable to a \textit{white-box} censor, provided that the censor has gained access to the weights of the model deployed.
We also evaluate the performance overhead of \sysname{} (\S~\ref{sec:eval-perf}), \ie the latency and bandwidth that it adds over our baselines, and its computational cost (\S~\ref{sec:eval-res}), \ie whether a generative obfuscator can keep pace with traffic in real time on commodity hardware without a GPU.

Overall, we see the schedule-driven architecture as the \textit{next generation} of tunnel obfuscators, and we present \sysname{} as an early step toward making such obfuscators practical and deployable at scale.
In \S~\ref{sec:discussion}, we discuss a few significant limitations of \sysname{} that remain open for future research.
Our specific instantiation of the schedule-driven architecture is quite unlikely to be the best one.
Yet, by detailing the rationale behind the design decisions that we made for \sysname{}, we hope that readers and other developers may find them useful in their own designs.

\section{Background and Related Work}
\label{sec:background}

\textbf{Detecting encrypted tunnels.}
Our threat model considers censors that actively detect and block the use of encrypted tunneling tools, such as VPNs and encrypted proxies.
Besides the information that they deem unwanted, censors increasingly target the encrypted tunneling tools that users run to circumvent censorship, in order to enforce their censorship policies~\cite{tschantz2016sok,xuebridging2024}.
In their standard forms, most tunneling protocols send headers or handshakes in plaintext, so a censor can detect these protocols easily with deep packet inspection~\cite{winter2012tor,xue2022openvpn}.
Such threats are not hypothetical.
GFW has targeted encrypted tunnels for more than a decade~\cite{winter2012tor,ensafi2015probing}, and other countries increasingly follow suit.
Iran, for example, blocked tunneling protocols as it escalated censorship during the 2022 protests~\cite{ooni2022iran,ioda2026iran}.
Russia similarly began blocking tunneling protocols in 2021 with TSPU~\cite{roskomsvoboda2023,xue2022tspu}.
More recently, the EU and the UK have debated restricting the use of tunneling tools because they let users bypass age verification~\cite{eprs2026vpnbriefing,uklords2026vpnamendment,ukcommons2026vpnreversal}.
Obfuscating a tunnel, \ie hiding the characteristics of tunneling, is therefore no longer a nice-to-have privacy feature, but close to a precondition for encrypted tunnels to work \textit{at all} in some networks, while the rest are visibly moving in the same direction.

\textbf{Obfuscating tunnel traffic.}
Circumvention tools obfuscate their tunnels with three classic strategies: (1) randomization, which encrypts every byte so that the traffic looks like random data~\cite{winter2013scramblesuit,obfs4spec,shadowsocksorg}; (2) mimicry, which shapes the traffic to imitate an allowed protocol~\cite{moghaddam2012skypemorph,weinberg2012stegotorus,dyer2013fte,dyer2015marionette}; and (3) tunneling, which carries the traffic inside a real session of an allowed application (\eg a video call)~\cite{fifield2015fronting,mcpherson2016covertcast,barradas2017deltashaper,barradas2020protozoa}.
The responses of censors to these obfuscators began a decade-long arms race that continues today~\cite{tschantz2016sok,xuebridging2024}.
Censors look for traits of non-standard implementations, such as rare cipher suites~\cite{winter2012tor,frolov2019tlscensorship} or a behavior that a mimic fails to reproduce~\cite{houmansadr2013parrot}, and obfuscators in turn imitate mainstream implementations (\eg the cipher suites of popular browsers~\cite{frolov2019tlscensorship}).
Censors actively probe suspected servers to see if they run an obfuscator~\cite{ensafi2015probing,shadowsocks2020imc,frolov2020probe}, and obfuscators counter with defenses against active probing~\cite{winter2013scramblesuit,obfs4spec}.
Censors deploy statistical tests that flag fully encrypted traffic such as Shadowsocks~\cite{wang2015seeing,shadowsocks2020imc,wu2023gfw}, and in response circumvention tools have moved to TLS-based tunnels.

Today, the most popular circumvention tunnels (\eg Trojan, VLESS, NaiveProxy~\cite{trojangithub,xraygithub,naiveproxygithub}) have converged on TLS.
These tunnels rely on the ubiquity of TLS, as blocking TLS outright would incur high collateral damage~\cite{frolov2019tlscensorship}.
The GFW nonetheless blocked TLS-based tunnels in 2022~\cite{gfwreport2022tlsblocking}, and recent fingerprinting attacks also focus on distinguishing TLS-based tunnels from normal TLS by their traffic shapes (sizes, timing and directions)~\cite{xue2024encapsulated,xue2025timing,ramesh2024calculatency,ali2026beyondrtt,censorkl,trojankiller}.

\textbf{Generative models of traffic.}
Generative models of network traffic have mostly been built to synthesize traces for testing or augmenting datasets, using GANs~\cite{lin2020doppelganger,yin2022netshare}, diffusion models~\cite{jiang2024netdiffusion} or state-space sequence models~\cite{chu2026netssm}.
These models generate whole traces offline and are judged by how useful the traces are for downstream tasks such as training classifiers, so none of them has to run inside a live connection or withstand a censor.
Closer to our setting, Surakav~\cite{gong2022surakav} defends Tor against website fingerprinting by shaping each page load to a burst sequence sampled from a GAN.
Voiceover~\cite{voiceover} obfuscates a covert channel inside a voice call by restricting it to the speaking patterns that a GAN learned from real two-person conversations.

\section{Why Yet Another Obfuscator?}
\label{sec:limits}

Today, there are a few dozen circumvention tunnel obfuscators that are deployed and operational (Appendix~\ref{app:survey} lists 30).
\textit{Why, then, propose yet another tunnel traffic obfuscator?}
To see this, we first make a distinction between two types of obfuscation architectures.
An \textit{application-driven} obfuscator (Fig.~\ref{fig:driven}, top) acts on the \texttt{write()} of the application layer (\eg a browser), each of which triggers its obfuscation primitives that may pad, split, or delay the data.
Importantly, the obfuscator does not act in the absence of an application \texttt{write()}, so it is idle whenever the application is idle.
In contrast, a \textit{schedule-driven} obfuscator (Fig.~\ref{fig:driven}, bottom) follows a schedule of its own, which dictates the size and the time of every record it sends.
Application data waits in a buffer, and whenever the schedule calls for a record, the obfuscator takes whatever data the buffer holds, up to the size of the record, and pads the rest.
Such a schedule-driven obfuscator can therefore send before any application data exists, continue while the application is idle, and hold data back while the schedule is quiet.

\begin{figure}[t]
\centering
\resizebox{\columnwidth}{!}{\includegraphics{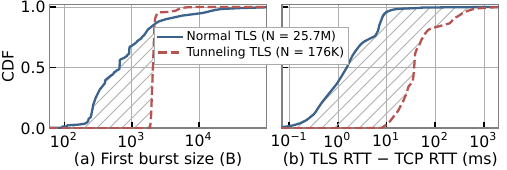}}
\caption{Distributional gaps that separate a TLS-based tunnel from \textit{normal} TLS, as described by prior fingerprinting attacks~\cite{xue2024encapsulated, xue2025timing, ali2026beyondrtt}.}
\label{fig:gaps}
\end{figure}

Application-driven obfuscators have certain advantages, the most significant of which is that they are easy to design and implement---a designer defines obfuscation primitives (\eg padding and delay) and triggers them on each application \texttt{write()}.
This is why nearly \textit{every} deployed obfuscator that we survey in Table~\ref{tab:survey} follows the application-driven design.
Yet, this design has a significant limitation, in that it can obfuscate \textit{in one direction only} (larger size and longer delay).
Consider two recent fingerprinting attacks on tunnels.
Xue et al.\ found that the size of the first few bursts after the TLS handshake~\cite{xue2024encapsulated}, as well as the apparent delays between those bursts~\cite{xue2025timing}, are highly effective fingerprints that separate normal TLS from TLS-based tunneling traffic, because larger bursts and longer round-trip delays are inherent to tunneling, as the distributions in Fig.~\ref{fig:gaps} show.
\textbf{These distributional gaps cannot be closed by an application-driven obfuscator}, as padding and delay driven by application events can only make burst size and delay larger, not smaller.
For example, Xue et al.~\cite{xue2025timing} found that the timing obfuscation of \texttt{obfs4}, an application-driven obfuscator, only made the timing fingerprint \textit{worse}, not better, as its delays make application-layer round trips appear even longer.
Essentially, application-driven obfuscation is inherently unable to close the distributional gap where the tunnel sits to the right of the traffic it wants to blend into.

These are only two examples, but they show that the inflexibility of application-driven obfuscation limits how effective and sustainable tunnel obfuscation can be.
For this reason, practitioners have been calling for schedule-driven designs for years~\cite{fifield2017insideout,fifield2023challenges,winter2019obfs5,xue2025timing}.
However, while the case for schedule-driven obfuscation is clear, designing and implementing such an obfuscator raises several non-trivial challenges.
First and foremost, what shape should its traffic take, and \textbf{what makes a shape good?}
Next, once we know the shape we are after, \textbf{how can we capture it?}
And finally, \textbf{how can that shape be reproduced at run time?}
These challenges are unique to schedule-driven obfuscation.
For example, existing application-driven obfuscators implicitly accept whatever shape results from the application traffic and their own obfuscation primitives.
A schedule-driven obfuscator, on the other hand, has no such default, so its designers must confront the question of what makes a shape good.

The rest of this paper describes our approach toward what we believe to be a practical and deployable schedule-driven obfuscator, which we tentatively name \sysname.
\S\S~\ref{sec:dataset}, \ref{sec:model} and~\ref{sec:system} address the three challenges above in turn.
We see schedule-driven designs as the \textit{next generation} of obfuscators, and with \sysname as an early step, we hope to turn the community's attention to their design and development.

\section{Defining Normal: A Campus Reference}
\label{sec:dataset}

\textit{What is a good schedule for a schedule-driven obfuscator?}
The arms race between censors and circumventors has shown that, for obfuscation to be sustainable in the long term, it is not enough to deviate from known tunnel fingerprints~\cite{xue2022openvpn,xue2024encapsulated,hanlon2024vpn,ramesh2024calculatency,xue2025timing,ali2026beyondrtt}.
The obfuscated traffic needs to blend into what the censor considers to be ``normal''.
Deviating from known fingerprints assumes a blocklisting censor that looks for those fingerprints, whereas blending into normal assumes a allowlist censor that scrutinizes anything statistically abnormal. Censors today are increasingly moving towards whitelisting~\cite{chen2026iranallowlist,wu2023gfw}.
A schedule-driven obfuscator should therefore strive to emulate a ``normal'' shape.
However, what is considered normal on the Internet is neither unique nor static, but depends on the visibility of the targeted censor (\eg a corporate firewall vs. a nation-state censor), on the composition of the monitored network (\eg data center vs. residential, how many users share one NAT), and on the cover protocol (\eg HTTPS vs. SMTP over TLS), and it changes as these factors do.
A meaningful definition of \textit{normal} for an obfuscation tool might therefore be the set of background traffic that is \textit{comparable} to the obfuscation traffic, \ie that the censor cannot distinguish from it through other means (\eg by IP address or protocol parser) and thus considers equivalent.

Following this reasoning, an obfuscator that fixes a static notion of ``normal'' cannot be practical.
\sysname therefore takes a modular design, in which the \textit{reference} for normal varies from deployment to deployment.
Users need to declare a target threat model and define what is \textit{comparable} to their obfuscation traffic from the perspective of the censor in that model.
For example, a user might collect (1) traffic to a particular website, if the SNI of the tunnel's cover is set to that website; (2) traffic from other users behind the same NAT, if the censor sits outside the NAT; or (3) traffic behind a gateway, if the censor sits outside the gateway and applies one classifier to all the traffic it sees there.

\subsection{The Campus Gateway Dataset}
\label{sec:dataset-data}

In this paper, we consider traffic collected at a campus gateway to be our reference for normal.
Specifically, on one selected core switch managed by our institution's IT department, we configured port mirroring to send a copy of sampled traffic (sampled at flow level, so both directions are selected/dropped together) to our \textit{Capture Point}.
The capture point runs Zeek~\cite{zeek} with a custom script and filters TCP connections on port 443.
For every connection, the script logs the times, sizes, SEQ/ACK, and flags of every TCP packet; for every TLS record, the script logs its content type, its record length, and its offset in the byte stream. The script then discards the raw payload.
A connection enters the dataset only if it passes three sanity checks: \emph{handshake integrity} (a full TCP handshake and a completed TLS handshake), \emph{stream integrity} (application data in both directions, no gaps in SEQ sequences that would indicate a packet loss) and \emph{length} (at least three application-data records and at most 500,000 packets).

\begin{figure}[t]
\centering
\resizebox{\columnwidth}{!}{\includegraphics{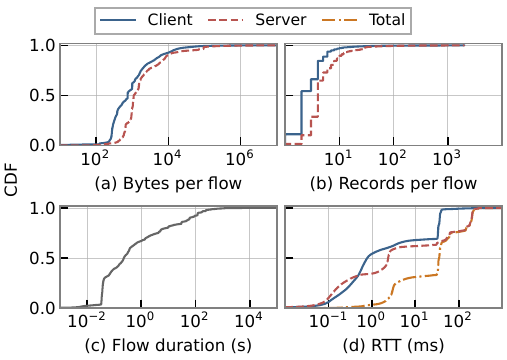}}
\caption{Characterization of the campus reference dataset.}
\label{fig:dataset}
\end{figure}

Our capture covered 33 days.
Of the 45.5\,M connections logged, the three sanity checks remove 36.9\,\%, 3.7\,\% and 2.3\,\%, respectively.
The remaining 26.0\,M connections (57.1\,\%) together hold 298.8M application-data records.
Fig.~\ref{fig:dataset} shows four statistical distributions of the dataset, in terms of the size and timing characteristics broken down into both directions.
The flows come from 17,954 client addresses and name 29,017 distinct servers in their SNI.
A breakdown of the server names by their service type can be found in Appendix~\ref{app:sni}.
We split the dataset once, at random with a fixed seed, into a training part of 90\,\% (23.4\,M flows) and a validation and a test part of 5\,\% each (1.3M) and keep the three parts fixed throughout.

For the rest of this paper, this dataset is the reference ``normal'' that \sysname aims to emulate, and all our evaluations are relative to it.
While we think that the dataset has the scale and heterogeneity to serve as a reference, we do not argue that it is the right one for every deployment, nor does it need to be.
The goal of \sysname is to take whatever traffic a user considers \textit{normal} and \textit{comparable}, learn from it, and generate new flows that are hard for the censor to isolate from that reference.

\section{Learning Normal: A Sequence Model}
\label{sec:model}
\begin{figure*}[t]
\centering
\resizebox{\textwidth}{!}{\includegraphics{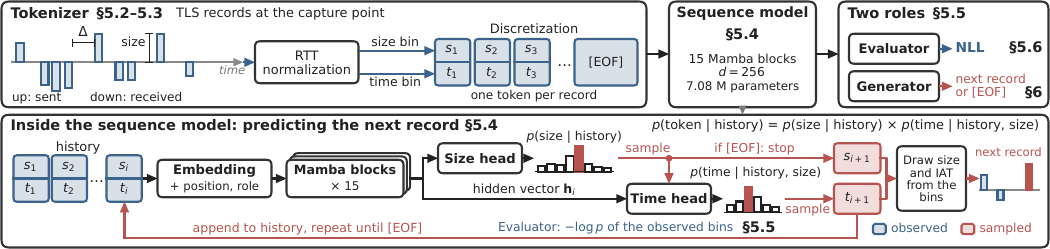}}
\caption{Overview of the traffic model, with the section that describes each part. Records at the capture point become tokens of a size bin and a time bin (blue). The model scores a flow by its NLL, or generates records by sampling the size first and then the time given the size (red).}
\label{fig:arch}
\end{figure*}

\subsection{Why a Generative Sequence Model}
\label{sec:model-why}

An effective tunnel obfuscator requires a source of new flows that a censor cannot distinguish from the flows of the reference.
Prior approaches model traffic as a low-order Markov chain, which draws each record either independently or conditioned on the preceding one or two records~\cite{moghaddam2012skypemorph,censorkl}.
However, the records of most network protocols, including TLS, do \textit{not} form a low-order Markov chain: a response, for example, depends on the request that initiated it, which may lie many records earlier.
Simplifying this dependence into a low-order model loses fidelity, while capturing it with a higher order leads to an explosion of contexts.
With a vocabulary similar to that of our sequence model (3,499 tokens), a mere order-2 Markov model already has about $4.3\times10^{10}$ parameters---two orders of magnitude more than the size of our dataset.
At the other extreme, prior work uses a replay pool to send pre-recorded sizes and times from real sessions~\cite{weinberg2012stegotorus}, which amounts to a Markov model whose order equals the length of a flow.
Besides the logistics of storing and maintaining the pool, this approach is fundamentally limited in its diversity.
Normal traffic consists of a small set of common shapes and a long tail of rare ones, and a pool holds only a sample of that tail.
The shapes in the sample make up the entire tail of the replayed flows, so these shapes recur several times as often as in ordinary traffic, and a censor with access to the pool can match them.

\textbf{A sequence model} addresses these problems.
Instead of storing a separate distribution for every context, it learns one function that updates a state with each record and predicts the next record from that state.
The number of parameters thus depends on this function and not on the number of contexts, so it does not grow with the length of the history (our model has 7.08\,M).
Histories that lead to similar states receive similar predictions, so what the model learns from one flow also applies to others~\cite{bengio2003neural}.
Unlike a replay pool, the model does not draw from a fixed set of recorded flows, but instead composes each new flow one record at a time. Thus it is not limited to the shapes it has seen, and it can condition each record on what the endpoint has actually received.
A sequence model thus differs from both extremes---unlike a low-order Markov model, it can represent dependence across a whole flow, and unlike a replay pool, it assigns probability to flows that it has never seen.

\subsection{Representing a Flow}
\label{sec:model-repr}

Fig.~\ref{fig:arch} shows an overview of our traffic model. 
An \sysname{} schedule---the shape of a flow---is represented at the granularity of TLS records.
Specifically, a flow is the sequence $f=(r_1,\dots,r_N)$ of the application-data records, in the order in which their first bytes pass the capture point.
Each record $r_i=(d_i,b_i,\tau_i^{\mathrm{s}},\tau_i^{\mathrm{e}})$ carries its direction $d_i$ ($d_i\in\{-1,\ +1\}$, +1 means client-to-server, -1 means server-to-client), its size $b_i$, and the times $\tau_i^{\mathrm{s}}$ and $\tau_i^{\mathrm{e}}$ at which the packets that hold its first and its last byte pass the capture point, which can differ (\eg when a TLS record is segmented into multiple TCP packets).

\textbf{RTT normalization.}
The raw timing of TLS records depends on the path as well as on the endpoints, and therefore needs to be normalized.
We measure the inter-arrival time (IAT) of $r_i$ from the end of the preceding record to the start of $r_i$, $\Delta_i=\tau_i^{\mathrm{s}}-\tau_{i-1}^{\mathrm{e}}$ (with $\Delta_1=0$), so that it excludes the transmission time of the preceding record.
An IAT then decomposes as
\begin{equation}
\Delta_i=D_{\mathrm{proc},i}+\rho_i+\epsilon_i,
\label{eq:decomp}
\end{equation}
where the processing time $D_{\mathrm{proc},i}$ is the time that the sender of $r_i$ takes before sending it, $\rho_i$ is the propagation delay of any round trip that $\Delta_i$ spans, and $\epsilon_i$ is jitter.
We want the model to learn $D_{\mathrm{proc},i}$ instead of the raw IAT $\Delta_i$, as the latter combines the propagation delays of the paths in the reference network with the behavior of its endpoints.

\begin{figure}[t]
\centering
\resizebox{\columnwidth}{!}{\includegraphics{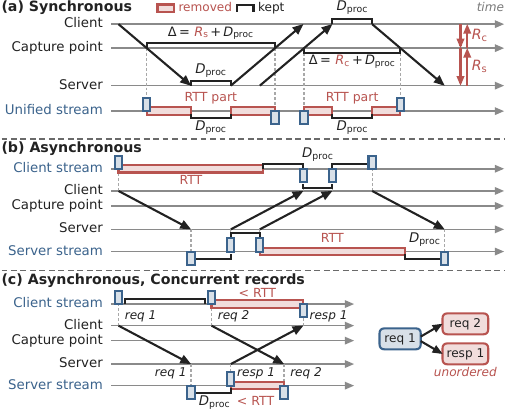}}
\caption{RTT normalization in the synchronous stream (a) and the asynchronous streams (b, c). For (c), the second request is sent before the first response arrives, which cross in transit and require special handling (\S~\ref{sec:model-dag}).}
\label{fig:norm}
\end{figure}

We therefore subtract\footnote{We subtract the round trip rather than divide by it, because division would also scale the processing time, which would then appear long on a short path and short on a long one.} an estimate $\hat\rho_i$ based on the TCP handshake RTT $R$, which is the sum of the round trips from the \textit{capture point} to the server ($R_{\mathrm{s}}$) and to the client ($R_{\mathrm{c}}$).
For two consecutive records in the same direction, $\Delta_i$ spans no round trip, so $\hat\rho_i=0$.
When the direction changes, assuming that $r_i$ answers the preceding record, $\Delta_i$ spans the round trip from the capture point to the sender of $r_i$ ($\hat\rho_i=R_{\mathrm{s}}$ if the server sent $r_i$, and $\hat\rho_i=R_{\mathrm{c}}$ if the client did).
The normalized IAT (Fig.~\ref{fig:norm}(a)) is therefore
\begin{equation}
\tilde\Delta_i=\max\bigl(0,\;\Delta_i-\hat\rho_i\bigr).
\label{eq:norm}
\end{equation}
The difference $\Delta_i-\hat\rho_i$ can be negative because of jitter and because of \emph{concurrent} records, where an endpoint may send a record before the preceding record of its peer has reached it, so that the two records cross in transit.
We address concurrent records in \S~\ref{sec:model-dag}.

\textbf{Discretization.}
Each record becomes a token $x_i=(s_i,t_i)=\bigl(Q_s(d_ib_i),\,Q_t(\tilde\Delta_i)\bigr)$, which pairs a size bin with a time bin from two quantizers that we fit on the training set.
The tokens of a flow form its \emph{synchronous} stream, which keeps the records of both endpoints in the order and \textit{on the clock of the capture point} (Fig.~\ref{fig:norm}(a)).
We model bins rather than the values themselves because both quantities concentrate probability mass on exact values.
A continuous distribution would need a density that grows without bound at such values.
A categorical distribution over bins, on the other hand, assigns probabilities in $[0,1]$, represents multimodal and heavy-tailed distributions without assuming their shape, and pools rare values into a small number of classes.

We use two different quantizers for size and IAT, as they carry information at different resolutions.
A size is set by the sender, and does not change along the network path, so $Q_s$ resolves sizes most finely where records are most frequent.
Thus, each frequent size receives a bin of its own, and the remaining sizes share bins of about equal probability mass.
$Q_s$ is fitted separately for the two directions.
An IAT, in contrast, carries information only at a resolution that depends on its magnitude.
Below a certain threshold (\eg 1\,ms), an IAT reflects network jitter rather than a decision of the application, whereas above the threshold, its order of magnitude distinguishes behaviors (\eg a response that required processing from a pause of the user).
$Q_t$ therefore consists of one bin below 1\,ms and, above it, bins of equal width in $\log\tilde\Delta$, with narrower bins around timer peaks.
For the synchronous stream, the fit yields $B_s=153$ size bins and $B_t=22$ time bins.
To turn a bin back into a value, the generator draws from the values that the bin received in training, which we store as empirical distributions $\pi_s(\cdot\mid s)$ over signed sizes and $\pi_t(\cdot\mid t)$ over normalized IATs.
Appendix~\ref{app:tokenizer} details the discretization of both sizes and IATs.

\textbf{End of flow.}
The length of a flow is part of the ``normal'' we are learning.
In fact, it varies over several orders of magnitude (Fig.~\ref{fig:dataset}(b)).
We therefore let the model learn to predict the end of a flow, rather than set the length of a generated flow by a rule outside the model.
Specifically, we tokenize at most the first $L=500$ records of a flow and stop once the sum of the normalized IATs exceeds $T=30$\,s.
A flow is \emph{complete} if a FIN or RST segment was observed and neither limit truncated it.
A complete flow receives a final token $\mathrm{[EOF]}=(B_s,B_t)$, which lies outside the range of both quantizers.
The generator then ends a flow when the model predicts [EOF].
Algorithm~\ref{alg:tokenize} in Appendix~\ref{app:tokenizer} summarizes the tokenization.

\subsection{Handling of Concurrent Records}
\label{sec:model-dag}

The synchronous stream (Fig.~\ref{fig:norm}(a)) assumes that the order in which records pass the capture point is the order in which they depend on each other.
This assumption fails for concurrent records, where the two endpoints send without waiting for each other (Fig.~\ref{fig:norm}(c)).
Two such records cross in transit, so each endpoint sees its own record before the record of its peer.
For example, with HTTP/2, a client may send its next request while the server is still answering the previous one.
The causal structure of such a flow (established by SEQ/ACK analysis) is therefore a directed acyclic graph (DAG) rather than a sequence.
As a result, the two endpoints hold divergent views of the flow's history.

Concurrent records such as those in Fig.~\ref{fig:norm}(c) are not a corner case that we could ignore by forcefully flattening the DAG into a sequence.
In the training part, the two endpoints order the records differently in 39\,\% of the flows.
Such crossings can make normalized IATs $\Delta_i-\hat\rho_i$ negative (\S~\ref{sec:model-repr}) in synchronous streams, which keep one linear order of the DAG (\ie for two adjacent records in opposite directions, $\Delta_i$ may be shorter than the actual RTT, because neither causally depends on the other).
More generally, learning sequence models of DAG-structured data is a known difficulty~\cite{vinyals2016order,you2018graphrnn}.

We therefore propose a second representation with two \emph{asynchronous} streams per flow---client stream and server stream---each of which keeps the records in the order and on the clock of its endpoint.
The stream $x^E$ of endpoint $E\in\{\mathrm{c},\mathrm{s}\}$ contains all $N$ records of the flow, with timing moved onto the clock of $E$ (assuming that the one-way delays between the capture point and $E$ are equal, $R_E/2$).
A record that $E$ sent left $E$ half a round trip before it passed the capture point, and a record that $E$ received reached $E$ half a round trip after, so we shift both times by $R_E/2$:
\begin{equation}
\alpha_i^E=\tau_i^{\mathrm{s}}-o_i^E\,R_E/2,\qquad \omega_i^E=\tau_i^{\mathrm{e}}-o_i^E\,R_E/2,
\label{eq:clock}
\end{equation}
where $\alpha_i^E$ and $\omega_i^E$ are the start and the end of $r_i$ on the clock of $E$, and $o_i^E \in \{-1,1\}$ (+1 means sent, -1 means received) marks the direction.
Within $x^E$, the records keep the order in which $E$ sent and received them, with the IAT on the clock of $E$ being $\Delta_i^E=\alpha_i^E-\omega_{i-1}^E$.
On this clock, a send followed by a receive spans the full round trip $R$, so the normalized IAT (Fig.~\ref{fig:norm}(b)) is
\begin{equation}
\tilde\Delta_i^E=\max\Bigl(0,\;\Delta_i^E-R\cdot\mathbf{1}\bigl[o_{i-1}^E=+1,\;o_i^E=-1\bigr]\Bigr).
\label{eq:norm2}
\end{equation}

\subsection{Sequence Model}
\label{sec:model-arch}

We train two models of the same structure, the \emph{synchronous model} on synchronous stream representation and the \emph{asynchronous model} on asynchronous stream representation.
Their backbone and training mostly follow standard practice for neural sequence models. We describe them at a high level here and refer the readers to Appendix~\ref{app:model} for the full details regarding the model architecture and their training.

Given a stream (synchronous or asynchronous) $x=(x_1,\dots,x_M)$, the sequence model predicts each token from its history, first the size, then the time bin given the size:
\begin{equation}
p_\theta(x_2,\dots,x_M\,|\,x_1)=\prod_{i=2}^{M}p_\theta(s_i\,|\,x_{<i})\,p_\theta(t_i\,|\,x_{<i},s_i).
\label{eq:factor}
\end{equation}
The size step and the timing step make no independence assumption.
We predict the size first because the size bin determines the direction of the record, and with it whether its IAT spans a round trip (\S~\ref{sec:model-repr} and\S~\ref{sec:model-dag}).
In addition, [EOF] is a size class, so the size decides whether a time is needed at all.
The NLL of token $x_i$, in nats, is
\begin{equation}
\ell_i=-\log p_\theta(s_i\,|\,x_{<i})-\log p_\theta(t_i\,|\,x_{<i},s_i).
\label{eq:nll}
\end{equation}

\textbf{Architecture.}
The model maps the two bin indices of a token, scaled to $[0,1]$, linearly to a vector of $d=256$ dimensions and adds a sinusoidal position code~\cite{vaswani2017attention}.
The asynchronous model also adds, at every position, a learned \emph{role vector} $\mathbf{r}_E$, which tells the model whether the stream belongs to the client or to the server.
The async model can thus serve both roles by learning what the two roles have in common from the streams of both, while the role vector lets the predictions differ where the roles differ.
The backbone, a stack of Mamba blocks~\cite{gu2024mamba}, summarizes the history $x_{\le i}$ in a hidden vector $\mathbf{h}_i$ and, like an actual endpoint, never uses the records that follow position $i$ (\ie only looks backward).
Mamba is a selective state-space model: the update of its state depends on the current input, so at every token the model can decide how much of its state to keep and how much of the input to write into it.
Its state also has a fixed size, so the cost of generating a record does not grow with the length of the flow, and training still processes all positions of a stream in parallel.
Two output heads compute the two factors of Eq.~\eqref{eq:factor}.
The \emph{size head} is a linear layer followed by a softmax over the $B_s+1$ size classes ([EOF] included).
The \emph{time head} receives an embedding of the size bin in addition to $\mathbf{h}_i$ and applies one hidden layer before a softmax over the $B_t+1$ time classes.

\textbf{Training.}
Training minimizes the mean of $\ell_i$ over every position $i\ge2$ of every training stream (Eq.~\eqref{eq:app-loss}), with the time head conditioned on the observed size bin (teacher forcing).
In an asynchronous stream, these positions also include the records that the owner of the stream received, so the model also learns to predict whether and when the peer sends next.
An endpoint uses these predictions to decide whether to wait for its peer and for how long (\S~\ref{sec:system-async}).
We train the asynchronous model on the 46.4\,M asynchronous streams of the training part (507\,M tokens) and the synchronous model on the 23.2\,M synchronous streams (253\,M tokens), each for 20 epochs, and keep the parameters with the lowest validation loss.

\begin{figure}[t]
\centering
\resizebox{\columnwidth}{!}{\includegraphics{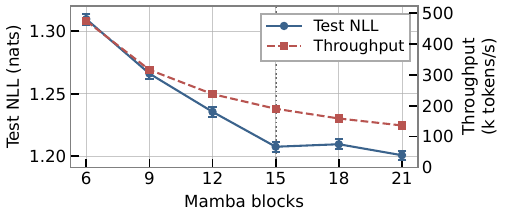}}
\caption{Test NLL and throughput of models across different model scale.}
\label{fig:scale}
\end{figure}

\begin{figure*}[t]
\centering
\resizebox{\textwidth}{!}{\includegraphics{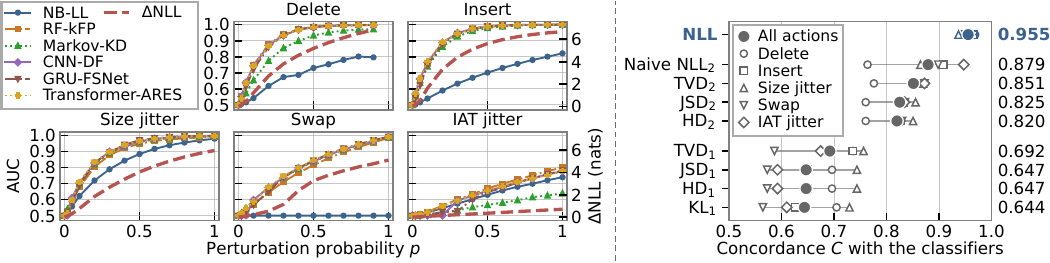}}
\caption{Left: the AUC of each of the six classifiers (left axis) and the change in the median flow NLL ($\Delta$NLL, right axis) against the perturbation probability $p$, per action. Right: the concordance $C$ of each distance metric with the distinguishability $D$ over pairs of conditions (large dots, values on the right), and per action over the comparable pairs.}
\label{fig:conc}
\end{figure*}

\textbf{Model scale.}
The depth of the backbone trades the accuracy of the model against its throughput (Fig.~\ref{fig:scale}).
Among asynchronous models of 6 to 21 Mamba blocks (Appendix~\ref{app:model}), the test NLL decreases from 1.310 at 6 blocks to 1.208 at 15 blocks and changes little beyond (1.201 at 21 blocks), while the scoring throughput on one GPU falls from 475\,k tokens per second at 6 blocks to 190\,k at 15 and 136\,k at 21.
We choose 15 blocks (7.08\,M parameters), which achieve 94\,\% of the NLL improvement from 6 to 21 blocks at 1.4 times the throughput of 21 blocks.

\subsection{Evaluator and Generator}
\label{sec:model-use}

The trained sequence model can serve either as an \textit{Evaluator} or a \textit{Generator}.
As an \textbf{\textit{Evaluator}}, it scores a flow $f$ by the mean of the NLL of Eq.~\eqref{eq:nll} over a set $S(f)$ of scored positions,
\begin{equation}
\mathrm{NLL}(f)=\frac{1}{|S(f)|}\sum_{(E,i)\in S(f)}\ell_i^E,
\label{eq:flownll}
\end{equation}
which we call the NLL of the flow.
For the synchronous model, $S(f)$ contains every position $i\ge2$ of the synchronous stream, including the [EOF] of a complete flow.
For the asynchronous model, $S(f)$ contains the positions $i\ge2$ of both streams whose record the owner of the stream sent ($o_i^E=+1$), together with the [EOF] of each stream of a complete flow.
A received record reflects a decision of the peer, so the asynchronous score counts it in the stream of the peer, where it is a sent record, and uses it only as context in the stream of its receiver.
Scoring [EOF] raises the NLL of a flow that ends where normal flows continue.

When used as a \textbf{\textit{Generator}}, it samples a stream from the model one token at a time, at temperature one, starting from a given first token (its opener, \S~\ref{sec:system-history}).
Each step samples a size bin, which ends the stream if it is [EOF], and otherwise samples a time bin given the size bin and draws a signed size from $\pi_s(\cdot\mid s)$ and a normalized IAT from $\pi_t(\cdot\mid t)$.
With the synchronous model, the generator samples all records of a flow before the first record is sent, as one plan that both endpoints follow (Algorithm~\ref{alg:sync}).
With the asynchronous model, each endpoint instead samples its next token as the flow proceeds, given its own stream, which also contains the records that the endpoint has received (Algorithm~\ref{alg:async}).

\subsection{NLL as a Proxy for Distinguishability}
\label{sec:model-proxy}

The generator samples each record from the model, favoring records to which the model assigns a low NLL.
The goal of an obfuscator, however, is traffic that classifiers cannot separate from ``normal''.
While we do not claim that a low NLL and indistinguishability are equivalent, we ask whether the NLL \textit{tracks} distinguishability closely enough to serve as a loose proxy for it.
To answer this, we perturb normal flows in controlled ways and compare how much each perturbation raises the NLL, versus how well classifiers distinguish the perturbed flows from the originals.

Specifically, we perturb 100,000 flows of the test part with five record-level actions (deleting, inserting or swapping records, or rescaling their sizes or IATs), each applied to every record with a probability $p$ from 0.02 to 1, which yields 59 perturbation conditions $\gamma$.
Next, for each condition $\gamma$, we train six classifiers to separate the perturbed flows from the clean ones on the same representation that our model sees.
The classifiers are a histogram likelihood ratio (NB-LL~\cite{liberatore2006inferring}), a random forest on hand-crafted features (RF-kFP~\cite{hayes2016kfp}), Markov chains (Markov-KD~\cite{korczynski2014markov}), and convolutional, recurrent and Transformer networks (CNN-DF~\cite{sirinam2018deep}, GRU-FSNet~\cite{liu2019fsnet} and Transformer-ARES~\cite{deng2023ares}).
The \emph{distinguishability} $D(\gamma)$ of a condition $\gamma$ is the highest area under the ROC curve (AUC) that any of the six classifiers achieves (Fig.~\ref{fig:conc}, left).
We then compute several \textit{distance metrics} between the clean and the perturbed flows, including the difference in their median flow NLL (Eq.~\eqref{eq:flownll}) and a few distance scores common in related work (TVD, JSD, HD and KL)~\cite{censorkl,chu2026netssm}, computed over single tokens (subscript 1) or pairs of consecutive tokens (subscript 2).
A distance metric that an obfuscator could optimize in place of distinguishability should order the conditions as $D$ does, which we measure with Harrell's concordance index $C$~\cite{harrell1982evaluating}---the share of pairs of conditions that the metric and $D$ \textit{order the same way}:
\begin{equation}
C=\frac{n_{\mathrm{c}}+\tfrac12 n_{\mathrm{t}}}{n_{\mathrm{c}}+n_{\mathrm{d}}+n_{\mathrm{t}}},
\label{eq:cindex}
\end{equation}
where $n_{\mathrm{c}}$, $n_{\mathrm{d}}$ and $n_{\mathrm{t}}$ count the comparable pairs that the metric orders as $D$ does, in the opposite order, or as ties, respectively.
Appendix~\ref{app:validation} details the perturbation actions, the classifiers, the distance metrics and the concordance score.

As shown in Fig.~\ref{fig:conc} (right), the NLL orders the conditions as the best-performing classifiers do, with $C=0.955$, against 0.82--0.85 for the other distance metrics over pairs of tokens and 0.64--0.69 for those over single tokens.
The concordance rises with the history that each metric conditions on (none for the single-token distances, one record for the pair distances, and the entire flow for the NLL).
This trend suggests that the dependence between records---including dependence that spans more than two records---carries much of what makes these perturbations distinguishable from normal traffic, which motivated our use of a sequence model and its NLL as a proxy for distinguishability.

We think that a matching NLL distribution is a \textit{necessary} condition for indistinguishability; it is not a sufficient one, though, because the NLL evaluates one flow at a time and so cannot capture aggregate properties of a set of flows.
We further discuss such limitations in \S~\ref{sec:discussion}.

\section{\sysname: A Generative Tunnel Obfuscator}
\label{sec:system}
The sequence model of \S~\ref{sec:model} forms the core of \sysname, as it provides what matters most to a schedule-driven obfuscator---its schedule.
An obfuscator, however, is more than a model of traffic, and turning the model into an obfuscator raises two system design problems that existing obfuscators do not face.
The first is that the obfuscation is now \textit{decoupled} from the application.
The schedule decides when each record is sent and how large it is, whereas the application writes data at its own times.
The second is \emph{synchronization}: a schedule spans both directions, so the two endpoints must cooperate to realize it.
Yet they are separated by the Internet, whose delay and jitter give each endpoint a different view of what has already happened in the flow.
How, then, can the two endpoints infer the next token when they do not even agree on the history that precedes it?
Here, we present our approaches to the two issues, the trade-offs they involve, and the rationale behind them.

\begin{figure*}[t]
\centering
\includegraphics{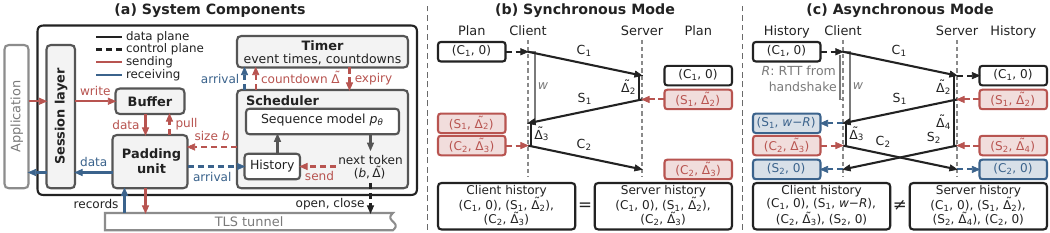}
\caption{(a) System components of \sysname. (b) and~(c) show the same records C$_1$, S$_1$ and C$_2$, under the two different operation modes. Red tokens are sampled by the model, blue tokens are computed from an arrival (Eq.~\eqref{eq:observe}). In Synchronous mode (b), each record waits its planned IAT after the preceding record at its sender, and a received record enters the history with its planned token (not actual timing), so the histories of two sides match. In Asynchronous mode (c), an endpoint adds the arrival to its history with the normalized IAT on its own clock, so the histories \textit{can} differ in their IATs.}
\label{fig:system}
\end{figure*}

\subsection{System Design}
\label{sec:system-arch}

Fig.~\ref{fig:system}a shows the five components of one \sysname endpoint, which interacts with both the application and the TLS tunnel.
The \emph{scheduler} obtains each next token from the sequence model given the \emph{history} of the flow at its endpoint (the tokens of the records that the endpoint has sent \textit{and} received).
The \emph{timer} records the time of every \emph{event} (a send or a receive) and runs the countdowns that the scheduler sets, interrupting the scheduler when one expires (\eg when a send is due).
The scheduler opens the TLS tunnel when it starts, and closes it when the next token is [EOF].

The \textit{buffer}, the \textit{padding unit} and the \textit{session layer} are the three components of \sysname{} that help it decouple the schedule from the application.
The application writes its data at its own pace into the \emph{buffer}, through an interface provided by the \emph{session layer}~\cite{fifield2019turbotunnelpost} (\eg KCP) that guarantees reliable delivery.
Rather than sending the data as it is written, \sysname \textit{pulls} data \textit{from} the buffer when the schedule calls for a record.
For a send of size $b$ after $\tilde\Delta$, the scheduler sets a countdown on the timer and, when the countdown expires, passes the size $b$ to the \emph{padding unit}.
The padding unit pulls from the buffer as much data as a record of size $b$ can carry and, if the buffer holds less, fills the rest of the record with padding before passing the record to the TLS tunnel.
When a record of the peer arrives, the padding unit removes the padding, passes the data through the session layer to the application and reports the arrival to the scheduler.
After every send and every arrival, the scheduler adds the record to the history and infers the next token.
When a schedule ends with [EOF] before the application's flow does, the scheduler opens a new TLS tunnel with a new schedule, and the session layer ensures the continuity of the application flow without interruption.

Decoupling costs bandwidth, as records that the schedule calls for while the application is idle carry only padding; decoupling also adds latency and limits the goodput, as data waits in the buffer for the schedule. We quantify such performance costs in \S~\ref{sec:eval-perf}.
One way to reduce the latency and raise the goodput is to maintain multiple concurrent TLS tunnels, each with its own schedule, so that together they offer more capacity than a single tunnel.
While concurrent tunnels are easy to run, they raise new questions outside the scope of this paper, such as how to obfuscate at the inter-flow level (\S~\ref{sec:discussion}).

\subsection{The Synchronization Problem}
\label{sec:system-history}

An \sysname{} schedule requires coordination between both endpoints, as the sequence model predicts each token conditioned on the past events of both sides---which creates a synchronization problem.
Sending is straightforward: after the \textit{opener} (the first records of the schedule), an endpoint sends each of its records once the time $\tilde\Delta$ of the record's token has passed since the endpoint's last event.
Receiving, on the other hand, is noisy.
The receiver knows only when the record arrived, and the normalized IAT that it computes \textit{on its own clock} will inevitably differ from the record's $\tilde\Delta$ by the change of the RTT (jitter) since the last RTT estimation.
The two endpoints will thus assign different normalized IATs to the same record.
Two endpoints that condition on different histories \textit{will} therefore diverge in their inference---they will be desynchronized.
As a result, they may predict conflicting next tokens (\eg both sends, leading to a race condition and further diverging the histories; or both receives, leading to a deadlock).
\sysname addresses this problem in two modes, which either prevent desynchronization (synchronous mode) or tolerate it and resolve the conflicts that it causes (asynchronous mode).

\subsubsection{Synchronous Mode}
\label{sec:system-sync}

The synchronous mode uses the model trained on the synchronous stream representation (\S~\ref{sec:model-repr}).
Here, both endpoints replace their observations with one \emph{plan}, a synchronous stream that begins with the opener and continues with records sampled from the synchronous model until [EOF].
An arrival still determines when the next record leaves, but it never enters the history, so the two endpoints hold the same history and cannot desynchronize.
Each endpoint samples the plan itself, from the same model, the same opener, and a random seed that the two endpoints share, which can be transmitted in-band, for example, in the random field of the TLS ClientHello~\cite{wustrow2011telex}.
Because the plan depends on no observation, an endpoint can sample it up to [EOF] before even sending its first record.
Each endpoint then walks through the plan, sending each of its own records the planned $\tilde\Delta$ after its last event and waiting without a deadline for each record of its peer.

Under the synchronous mode, the record-sending is strictly sequential, so a synchronous flow cannot deadlock, nor can a receive interrupt a pending send (no race condition).
However, it also produces no concurrent records, since no record leaves while a record of the peer is in transit, whereas the endpoints of 39\,\% of the training flows send concurrently (\S~\ref{sec:model-dag}). 
The plan also cannot respond to the network, since a delay on the path postpones the records that follow it but does not affect the inference of the model.
Algorithm~\ref{alg:sync} describes the operation of \sysname{} under synchronous mode.

\begin{algorithm}[t]
\caption{Synchronous mode}
\label{alg:sync}
\begin{algorithmic}[1]
\Require synchronous model $p_\theta$; opener; shared seed
\State $x\gets$ the opener \Comment{one plan per flow}
\Repeat\ sample $y\sim p_\theta(\cdot\mid x)$ and append $y$ to $x$
\Until{$y=\mathrm{[EOF]}$}
\Statex Each endpoint $E$ then walks through the plan $x$:
\For{each record $(\sigma_i,\tilde\Delta_i)$ of $x$}
  \If{$E$ sends record $i$} wait $\tilde\Delta_i$, then send it
  \Else\ wait for its arrival
  \EndIf
\EndFor
\State close the connection \Comment{[EOF]}
\end{algorithmic}
\end{algorithm}

\subsubsection{Asynchronous Mode}
\label{sec:system-async}

The asynchronous mode uses the model trained on the asynchronous stream representation (\S~\ref{sec:model-dag}).
Here, each endpoint keeps its observations and conditions on its own stream (its own sends and the arrivals that it observes), together with the role vector of its side (\S~\ref{sec:model-arch}).
A record that arrives at endpoint $E$ enters its stream with the normalized IAT on the clock of $E$,
\begin{equation}
\tilde\Delta_i^E=\max\bigl(0,\;w-R\cdot\mathbf{1}[o_{i-1}^E=+1]\bigr),
\label{eq:observe}
\end{equation}
where $w$ is the time that has passed since the last event, $o_{i-1}^E=+1$ if $E$ sent the preceding record, and $R$ is the RTT that $E$ measured from the TCP handshake.
Because each endpoint conditions on what it observes, the asynchronous mode responds to the network, and its two endpoints can send concurrently, so that their records cross in transit as in real flows (Fig.~\ref{fig:system}c).

Under the asynchronous mode, however, the two endpoints decide independently from histories that differ, leading to potential race conditions that need explicit handling.
First, the record of the peer may arrive while an endpoint counts down to a send that it sampled before the arrival (Fig.~\ref{fig:async}a).
The endpoint then cancels the send, adds the arrival to its history and samples again, so that every record it sends is a sample given its current stream.
Second, both endpoints may predict a receive, so that neither sends and the flow enters a deadlock (Fig.~\ref{fig:async}b).
To break such a deadlock, an endpoint that predicts a receive waits only until the deadline
\begin{equation}
\kappa+\alpha\,(F^E)^{-1}(\lambda)+R+\beta,
\label{eq:deadline}
\end{equation}
where $\kappa$ is the time of the last event at $E$ and $(F^E)^{-1}(\lambda)$ is the $\lambda$-quantile of the normalized IAT that the model predicts for the next received record.
The quantile makes the deadline short within a burst and long before a pause.
We draw $\lambda$ from $[0.90,0.95]$ for each deadline and use $\alpha=2$ and $\beta=10$\,ms, so that an endpoint stops waiting only when the record of its peer is well overdue under the model, which suggests that the peer is waiting as well.
When the deadline passes, the endpoint samples again from the time bins at or above $w$, which yields a send, an [EOF] or another receive with a new deadline.
Algorithm~\ref{alg:async} describes the operation of \sysname{} under the asynchronous mode.

\begin{algorithm}[t]
\caption{Asynchronous mode at endpoint $E$}
\label{alg:async}
\begin{algorithmic}[1]
\Require asynchronous model $p_\theta$; opener; RTT $R$
\If{$E$ has records in the opener} send them
\Else\ wait for an arrival
\EndIf
\State $x^E\gets$ the records so far;\; $\kappa\gets\mathrm{now}$;\; $k\gets0$
\Loop
  \State $w\gets\mathrm{now}-\kappa$ \Comment{time already waited}
  \State sample $y\sim p_\theta(\cdot\mid x^E)$ with time bins $\ge w$
  \Statex \hspace{\algorithmicindent}(no receive if $k>4$ or $w$ is in the largest bin)
  \If{$y=\mathrm{[EOF]}$} close the connection and \Return
  \EndIf
  \If{$y$ is a send of size $\sigma$ and IAT $\tilde\Delta$} $\kappa'\gets\kappa+\tilde\Delta$
  \Else\ $\kappa'\gets$ the deadline of Eq.~\eqref{eq:deadline}
  \EndIf
  \State wait until $\kappa'$, an arrival or the closing
  \If{the peer closed the connection} \Return
  \EndIf
  \If{a record arrived} append it as in Eq.~\eqref{eq:observe}
  \ElsIf{$y$ is a send} send and append $y$
  \Else\ $k\gets k+1$; \textbf{continue} \Comment{the deadline passed}
  \EndIf
  \State $\kappa\gets\mathrm{now}$;\; $k\gets0$
\EndLoop
\end{algorithmic}
\end{algorithm}

\subsection{Implementation}
\label{sec:system-implementation}

We implemented a prototype of \sysname in Golang.
Implementation details and the open-source repository can be found in Appendix~\ref{app:openscience}.

\section{Evaluation}
\label{sec:eval}

We evaluate \sysname along three axes, under various controlled network conditions.
The first is \emph{distinguishability}, \ie how well censors under different threat/capability models distinguish \sysname{} flows from normal flows, and how the RTT and the jitter of the network affect it (\S~\ref{sec:eval-quality}).
The second is \emph{overhead}, \ie the latency and the bandwidth that \sysname adds when it carries the data of the application inside the records of a schedule, compared with normal traffic that carries the same data (\S~\ref{sec:eval-perf}).
Finally, we evaluate \sysname{}'s \emph{computational cost}, \ie whether resources provided by reasonable hardware run \sysname{} fast enough (in particular, the inference step) to decide each record in real time (\S~\ref{sec:eval-res}).

\textbf{Setup.}
In every experiment, a pair of \sysname client and server exchange the records of a flow across a local testbed network, with controlled variation over network conditions (RTT $R=100$\,ms and jitter $b=0$ unless stated otherwise).
Each direction of the path delays a record by $R/2$ (optionally plus a jitter).
The capture point, where the censor observes the flow, lies at the midpoint of the path (\ie $R_{\mathrm{c}}=R_{\mathrm{s}}=R/2$), where we tokenize the traffic log into the synchronous stream and the two asynchronous streams described in \S~\ref{sec:model-repr} and~\ref{sec:model-dag}.
Both endpoints run \sysname with the 15-block models (\S~\ref{sec:model-arch}), which we trained on the training part of the campus reference.
We evaluate both the synchronous mode with the synchronous model and the asynchronous mode with the asynchronous model, and call the traffic that they generate \emph{sync} and \emph{async}.

\textbf{Baselines.}
We use the test part of the campus reference dataset as a \emph{Normal} baseline.
Additionally, we build \emph{Replay (sync)} and \emph{Replay (async)}, which play the records of the \emph{Normal} flows in synchronous and asynchronous mode, respectively, without the model, to isolate the effect of the \textit{schedule} on (in)distinguishability from that of the \textit{mode} and the network.
Unless noted otherwise, every experiment uses 2,000 flows per cell (\ie per class and per network setting).

\subsection{(In)Distinguishability}
\label{sec:eval-quality}

We evaluate how well censors can distinguish \sysname flows from normal flows.
While the NLL tracks distinguishability across perturbations (\S~\ref{sec:model-proxy}), here we test distinguishability directly, with classifiers that a censor could train to separate \sysname flows from normal flows.
\textbf{Caveat}: we caution readers that testing a security property (\eg indistinguishability) with attacks is hard, because immunity against specific classifiers does \textit{not} rule out exposure to classifiers that we did not test.
We therefore use classifiers of different types, in the hope that the best of these classifiers approximates the best one that a censor could build.

Our censors build on the six classifiers that span three types: marginal classifier (NB-LL, a histogram likelihood ratio), classical machine-learning classifiers (RF-kFP, a random forest on hand-crafted features, and Markov-KD, which fits Markov chains), and three deep neural networks (CNN-DF, GRU-FSNet and Transformer-ARES).
We evaluate \sysname under three threat models, which differ in the policy of the censor (blocking known tunnels vs. allowing only normal traffic) and in what the censor knows about \sysname:
\begin{itemize}[\setlength{\IEEElabelindent}{0pt}\setlength{\labelsep}{0.5em}]
\item A \emph{blocklisting censor} aims to block known tunneling traffic.
It trains the six classifiers on normal flows (40,000 flows of the test part) against flows of nine tunneling protocols (Trojan, VLESS, VMess, each over TLS, WebSocket/TLS and REALITY, with 4,445 flows each).
\item An \emph{allowlisting censor} knows that obfuscators exist, \sysname among them, and therefore allows only flows that look \textit{normal}.
Since it learns from normal flows alone, it uses one-class forms of above-mentioned classifiers, which we fit on 40,000 flows of the training part. 
\item A \emph{white-box censor} targets an \sysname deployment itself, holding not only its code, which is open source, but also the weights of the models that a deployment uses.
It can therefore generate flows of that deployment and train classifiers on them.
\end{itemize}

Table~\ref{tab:eval-censors} gives the AUC of every classifier and, for each  censor threat model, the highest AUC $D$ achieved by any classifier, along with the recall at a false positive rate of 0.6\% (the FPR used by GFW against fully encrypted tunnels as estimated by Wu et al.~\cite{wu2023gfw}).
Both \textit{async} and \textit{sync} reduce the blocklisting censor, whose classifiers separate tunneling flows completely, to little better than random guessing ($D=0.561$--$0.581$).
\sysname also provides a reasonable defense against an allowlisting censor, which separates tunnel flows well ($D=0.965$) but async and sync flows only weakly ($D$ of 0.642 and 0.605).
Against the white-box censor, however, $D$ rises from $0.642$ to $0.764$ for \textit{async} and similarly for \textit{sync}.
A deployment shares its weights only with its users, much as a Shadowsocks/obfs4 server shares its access key, and the threat models of these tools typically assume that a censor does not hold such a secret.
The white-box censor becomes relevant, though, when the weights of a deployment are public, for example when users who cannot capture enough traffic of their own network use one that someone with such access has trained and released.

\begin{table}[t]
\centering
\caption{AUC of every classifier at $R=100$\,ms, and $D$ (the best AUC) with that classifier's recall (\%) at 0.6\,\% false positives. R(async) and R(sync) are the replays.}
\label{tab:eval-censors}
\scriptsize
\setlength{\tabcolsep}{1.4pt}
\renewcommand{\arraystretch}{0.9}\setlength{\aboverulesep}{0.2ex}\setlength{\belowrulesep}{0.35ex}
\begin{tabular*}{\columnwidth}{@{\extracolsep{\fill}}c@{\hspace{1pt}}lcccccc@{}}
\toprule
& Classifier & Normal & Tunnel & R(async) & Async & R(sync) & Sync \\
\midrule
\multirow{7}{*}{\rotatebox[origin=c]{90}{\textbf{Blocklist}}}
& NB-LL & .501 & .989 & .501 & .490 & .498 & .453 \\
& RF-kFP & .501 & 1.000 & .515 & .581 & .526 & .546 \\
& Markov-KD & .494 & .998 & .488 & .550 & .495 & .505 \\
& CNN-DF & .501 & 1.000 & .504 & .493 & .496 & .561 \\
& GRU-FSNet & .498 & 1.000 & .497 & .472 & .505 & .483 \\
& Transformer-ARES & .500 & 1.000 & .510 & .510 & .513 & .557 \\
& $D$ (recall, \%) & \textbf{.501}(0.6) & \textbf{1.000}(100) & \textbf{.515}(0.8) & \textbf{.581}(1.8) & \textbf{.526}(2.4) & \textbf{.561}(1.6) \\
\midrule
\multirow{4}{*}{\rotatebox[origin=c]{90}{\textbf{Allowlist}}}
& IF-kFP & .496 & .831 & .497 & .526 & .497 & .530 \\
& Markov-NLL & .499 & .914 & .503 & .563 & .504 & .511 \\
& AE-FSNet & .498 & .965 & .523 & .642 & .516 & .605 \\
& $D$ (recall, \%) & \textbf{.499}(0.5) & \textbf{.965}(18.4) & \textbf{.523}(0.6) & \textbf{.642}(1.7) & \textbf{.516}(0.4) & \textbf{.605}(3.0) \\
\midrule
\multirow{7}{*}{\rotatebox[origin=c]{90}{\textbf{White-box}}}
& NB-LL & .505 & -- & .583 & .630 & .513 & .606 \\
& RF-kFP & .515 & -- & .654 & .744 & .709 & .838 \\
& Markov-KD & .493 & -- & .525 & .743 & .602 & .788 \\
& CNN-DF & .495 & -- & .520 & .764 & .648 & .813 \\
& GRU-FSNet & .477 & -- & .440 & .640 & .658 & .785 \\
& Transformer-ARES & .511 & -- & .599 & .737 & .513 & .756 \\
& $D$ (recall, \%) & \textbf{.515}(1.3) & -- & \textbf{.654}(3.5) & \textbf{.764}(7.8) & \textbf{.709}(2.9) & \textbf{.838}(5.2) \\
\bottomrule
\end{tabular*}

\end{table}

\subsubsection{Impact of RTT and Jitter}
\label{sec:eval-nll}

In deployment, \sysname runs over paths whose RTT and jitter differ from those of the paths in its training data, so we also measure how the RTT and the jitter of the path affect obfuscation fidelity.
We vary the RTT over 10--200\,ms, and also add to the delay a jitter drawn from a Laplace distribution with a scale $b$ of 0.1--10\,ms (as the Laplace distribution is a common model for Internet jitter~\cite{houmansadr2014rainbow,elices2013flowcorrelation}), and measure the distinguishability.

\begin{figure}[t]
\centering
\resizebox{\columnwidth}{!}{\includegraphics{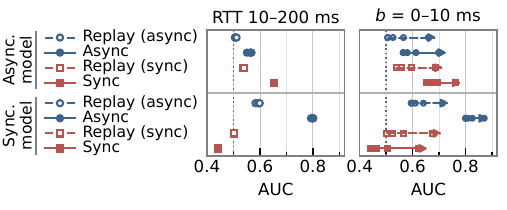}}
\caption{AUC of the flow NLL against normal flows, for each set and \{sync, async\} model, at different RTTs without jitter (left) and at $R=100$\,ms as the scale $b$ of the Laplace jitter grows along the axis (right).}
\label{fig:eval-nll}
\end{figure}

Fig.~\ref{fig:eval-nll} shows the AUC with which the flow NLL separates each set from a random sample of 50,000 normal flows of the test part.
Overall, the RTT leaves the AUCs of sync and Replay (sync) unchanged, because the synchronous mode sends each record its planned IAT after the last event of its sender, so a longer path delays the records without changing their normalized IATs.
The AUCs of async change by at most 0.02 between 10 and 200\,ms, as the RTT can change which records cross in transit and thus the order of the records in an endpoint's stream (\S~\ref{sec:system-async}).
Jitter, in contrast, raises every AUC once $b$ approaches 1\,ms, the width of the lowest time bin.
At $b=10$\,ms, the AUCs are 0.08--0.22 higher than without jitter under either the asynchronous or the synchronous model.
However, the baseline replays also rise as much as the generated sets, so the rise is an effect of the network rather than of the model, as the jitter changes the IATs that the capture point observes in real and generated flows alike.
Intuitively, a path with much more jitter than the training data therefore makes any flow \textit{less normal}, generated or not, which favors training the model on traffic of the network in which it is deployed.

\subsection{Performance Overhead}
\label{sec:eval-perf}

\begin{figure}[t]
\centering
\resizebox{\columnwidth}{!}{\includegraphics{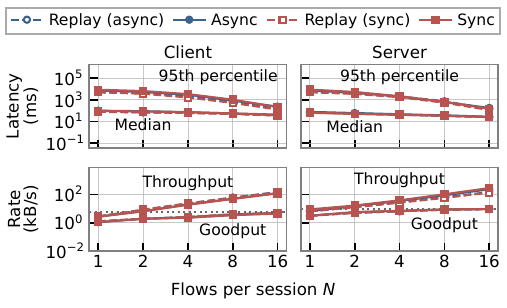}}
\caption{Sessions of $N$ flows of each set at $R=100$\,ms, per side: (top) the latency, median and 95th percentile over the sessions; (bottom) the median throughput/goodput, with the dotted line showing the write rate.}
\label{fig:eval-session}\label{fig:eval-session-lat}\label{fig:eval-session-rate}
\end{figure}

We now quantify the two costs of decoupling the schedule from the application (\S~\ref{sec:system-arch}), \ie the time that data waits in the buffer and the bandwidth that padding takes, for a single flow at different RTTs (Table~\ref{tab:eval-flow}) and for sessions of concurrent flows (Fig.~\ref{fig:eval-session}).

\begin{table}[t]
\centering
\caption{The median latency (Lat., ms) and the median throughput and goodput (Thr./Gp., kB/s) measured with one flow at each RTT, for the client (C) and the server (S).}
\label{tab:eval-flow}
\scriptsize
\setlength{\tabcolsep}{1.6pt}
\renewcommand{\arraystretch}{0.9}\setlength{\aboverulesep}{0.2ex}\setlength{\belowrulesep}{0.35ex}
\begin{tabular*}{\columnwidth}{@{\extracolsep{\fill}}llcccccccc@{}}
\toprule
 & & \multicolumn{2}{c}{$R=10$\,ms} & \multicolumn{2}{c}{50\,ms} & \multicolumn{2}{c}{100\,ms} & \multicolumn{2}{c}{200\,ms} \\
\cmidrule(lr){3-4}\cmidrule(lr){5-6}\cmidrule(lr){7-8}\cmidrule(l){9-10}
Set & & Lat. & Thr./Gp. & Lat. & Thr./Gp. & Lat. & Thr./Gp. & Lat. & Thr./Gp. \\
\midrule
Replay (async) & C & 24 & 12.1/2.8 & 50 & 5.2/1.8 & 78 & 2.8/1.3 & 131 & 1.5/0.8 \\
 & S & 88 & 7.8/2.4 & 55 & 10.4/3.5 & 70 & 6.4/3.1 & 114 & 3.7/2.8 \\
Async & C & 28 & 10.6/2.7 & 65 & 4.8/1.5 & 100 & 2.7/1.1 & 162 & 1.4/0.8 \\
 & S & 97 & 11.3/3.4 & 64 & 12.1/4.2 & 78 & 6.6/3.1 & 131 & 4.4/3.1 \\
Replay (sync) & C & 21 & 12.6/3.0 & 46 & 5.2/1.9 & 81 & 2.8/1.3 & 141 & 1.5/0.8 \\
 & S & 86 & 8.3/2.5 & 49 & 10.2/4.1 & 68 & 6.4/3.1 & 116 & 3.6/3.0 \\
Sync & C & 32 & 10.8/2.7 & 61 & 4.9/1.6 & 96 & 2.7/1.1 & 154 & 1.4/0.8 \\
 & S & 102 & 16.8/2.9 & 51 & 13.4/4.2 & 64 & 8.9/3.1 & 113 & 4.6/3.1 \\
\bottomrule
\end{tabular*}

\end{table}

\textbf{Setup}: At each endpoint, an application process writes into the buffer at Poisson times from the first to the last record of the flow.
Each write has the median record size of its side (177\,B for client and 314\,B for server), and each side writes at the median byte rate of the training part (5,378 and 9,906\,B/s).
As a baseline reference, the replays carry the same application-level writes, but on the schedules of real flows.
From this setup, we measure \textit{latency}, which is the mean time that the bytes written by the application spend in the buffer (without counting RTT).
Additionally, we measure \textit{throughput} and \textit{goodput} by the number of bytes sent on the wire (and the number of application bytes among them), divided by the duration of the flow.

As shown in Table~\ref{tab:eval-flow}, the latency grows with the RTT, from 28--32\,ms to 154--162\,ms with async and sync, and 21--24\,ms to 131--141\,ms for the replay baselines.
The latency grows because a write waits for the next record of its side, which often follows a reply that is a round trip away.
As expected, the throughput and the goodput fall as the RTT grows, from 10.6--16.8 and 2.7--3.4\,kB/s at 10\,ms to 1.4--4.6 and 0.8--3.1\,kB/s at 200\,ms, since a longer path spreads the records of a flow over more time.
For the median flow, the performance of async and sync tracks that of the replays.
Between the two modes, the latency of sync is 4--18\,ms lower than that of async, but sync pads more, as its server sends 5--35\,\% more bytes per second at the same goodput.

We caution against comparing the throughput of a single \sysname flow with that of an application-driven obfuscator.
An application-driven obfuscator sends whenever the application writes, so its throughput is bounded by the link and by congestion control, whereas \sysname sends only when its schedule calls for a record, so the throughput of one \sysname flow is bounded by that of a normal flow that its schedule imitates.
At 100\,ms, for example, a single flow carries only about a third of what the server's application writes, in sync, async, and the replay baselines alike (Table~\ref{tab:eval-flow}).
What \sysname can carry grows instead with the number of concurrent flows, $N$, which we measure in Fig.~\ref{fig:eval-session}.
With every increase of $N$, the latency falls and the goodput rises.
Concurrent flows, however, raise the inference cost (especially at a server that serves many clients) as well as obfuscation challenges at the inter-flow level, such as how many connections a client may keep open, which lie outside the scope of this paper (\S~\ref{sec:discussion}).

\subsection{Computational Cost}
\label{sec:eval-res}

\sysname runs a sequence model at its core, which makes it more resource-intensive than existing obfuscators.
The cost matters most in the asynchronous mode, because the synchronous mode can sample the whole plan of a flow in advance, whereas an asynchronous endpoint must decide its next record in real time after each record that it receives.
If such a decision takes longer than the IAT that it yields, the sender cannot realize the schedule faithfully.
We therefore estimate the computational cost of running \sysname{}.

\begin{figure}[t]
\centering
\resizebox{\columnwidth}{!}{\includegraphics{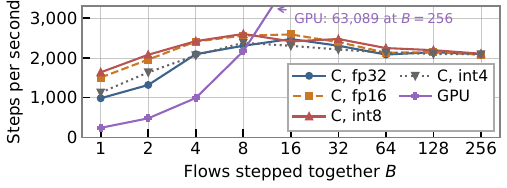}}
\caption{Model steps per second against the number of flows $B$ stepped together, on one CPU core at four weight precisions vs. on the GPU.}
\label{fig:eval-steps}
\end{figure}

For inference, we reimplemented the step of the model in C, with the weights stored in 32-, 16-, 8- or 4-bit numbers (fp32, fp16, int8 or int4) and with the steps of $B$ flows computed together (batching).
We measure the C step on one physical core of an Intel Xeon Gold 6459C (2\,MB of L2 and 60\,MB of L3 cache), where it uses AVX-512 instructions, and compare with an NVIDIA RTX 5090 that runs the Mamba CUDA kernels.
Table~\ref{tab:eval-cpu} and Fig.~\ref{fig:eval-steps} show the results.
On the Xeon core, a decision (one step and the sampling of a token) takes 1.15--1.19\,ms with the synchronous/asynchronous model at fp32, and 0.65--0.75\,ms with int8 weights.
The int8 weights keep the behavior of the model, since 99.8\,\% of its draws are identical to those of the GPU and the median flow NLL changes only by 0.004 (less than it changes between two seeds), whereas int4 weights change a quarter of the draws.
The GPU, for comparison, takes about 4\,ms per step at every $B$, because its step consists of many small kernel launches, so it is slower than one CPU core \textit{for a single flow} and faster only for many flows at once (63,000 steps per second at $B=256$).
Batching also raises the throughput of one CPU core to 2,382--2,604 steps per second at $B=8$ or 16 (Fig.~\ref{fig:eval-steps}), although each batched step takes longer.

\begin{table}[t]
\centering
\caption{One model step on the GPU and on one CPU core (C at four weight precisions): weights, mean time at $B=1$, best steps per second over $B$, and agreement (measured in mean KL) with the CUDA kernels.}
\label{tab:eval-cpu}
\footnotesize
\setlength{\tabcolsep}{3.5pt}
\begin{tabular}{@{}lrrrrr@{}}
\toprule
Engine & MB & ms & Steps/s ($B$) & KL & Same (\%) \\
\midrule
GPU, fp32 & 28.3 & 4.15 & 63,089 (256) & 0 & 100.0 \\
C, fp32 & 28.9 & 1.02 & 2,475 (16) & 0 & 100.0 \\
C, fp16 & 15.4 & 0.66 & 2,591 (16) & $4\cdot10^{-7}$ & 99.9 \\
C, int8 & 8.8 & 0.61 & 2,604 (8) & $3\cdot10^{-4}$ & 99.8 \\
C, int4 & 5.4 & 0.89 & 2,382 (8) & 0.39 & 75.4 \\
\bottomrule
\end{tabular}

\end{table}

We estimate the traffic that one core can shape by multiplying its steps-per-second by the mean number of bytes-per-record at $R=100$\,ms (bidirectionally 1,905--3,473\,B for sync and async), which gives about 23\,Mbit/s for sync mode with one decision at a time and 40\,Mbit/s with batching, and 37 and 72\,Mbit/s for async.
We therefore believe that \sysname runs in real time on a single CPU core without a GPU, with 8.8\,MB of int8 weights.

\section{Discussion and Conclusion}
\label{sec:discussion}

\textbf{Limitations.}
Our evaluation shows that \sysname, though much harder to distinguish than the tunneling protocols we tested, is still far from being completely indistinguishable (\S~\ref{sec:eval-quality}).
Part of the remaining distance does not come from the sequence model, since even \textit{normal} flows, replayed through either mode without the model, remain distinguishable to a white-box censor.
One reason is that both modes of \sysname rely on mechanisms that are designed rather than learned from normal traffic, such as the strictly sequential sending in the sync mode, which never lets records cross in transit, and the deadline mechanism with which the async mode breaks a deadlock.
Each such mechanism may become a fingerprint, because the code is open source and normal traffic has no reason to follow it.
For example, if the deadline is a fixed value, or follows a fixed rule even with some randomness, a censor that knows the rule may recognize the behavior with enough observations.
We see, however, no good way to avoid such mechanisms, because of the synchronization problem we described in \S~\ref{sec:system-history}.

\textbf{Beyond obfuscating a single flow.}
We focused on single flows because, by a simple reduction, a flow ensemble can be indistinguishable only if each of its flows is indistinguishable.
A censor that examines flow ensembles can apply any per-flow classifier to each flow of an ensemble, so a classifier that separates single \sysname flows from normal flows also separates the ensembles that contain them.
If \sysname{} turns out to be useful, a censor then has to look for patterns across the flows of a host, such as \textit{how many flows are open at once} and \textit{how long each lasts}.
We consider it \textit{progress} to force a censor to look across flows, yet these inter-flow patterns remain to be obfuscated.
Implementing this work also raised a question that we had not anticipated: \textit{when should \sysname run?}
Running it all the time spends bandwidth on padding and computation on inference while the application is idle.
Opening a flow only when the application has data to send, on the other hand, as we do now, lets the application decide when flows exist, which makes the obfuscator application-driven once more, at the inter-flow level.

Finally, some properties of traffic belong to both levels, such as the duration of a flow---set by its schedule but also determining inter-flow concurrency. We think that intra-flow and inter-flow obfuscation should therefore not be designed in isolation.
One possible approach is a hierarchical obfuscator, where a second sequence model decides when each flow opens and how long it lasts, and the model of \S~\ref{sec:model} generates the records within each flow, conditioned on the duration, as in length-controlled generation~\cite{kikuchi2016controlling}.

\textbf{Who trains the model?}
During evaluation, we asked ourselves whether the weights of a deployment should be shared only with its intended users, as a Shadowsocks or \texttt{obfs4} server does with its secret tokens~\cite{ss2022spec,obfs4spec}.
Secrecy of the weights raises the detection difficulty significantly, but also prevents them from being shared publicly, so that each deployment must train its own model, and how distinguishable its flows are, which depends on the weights, can be evaluated only by those who hold them.
Who trains the model also decides whose normal traffic it imitates.
A model of a whole gateway makes each flow typical of that gateway, but a host whose every flow is drawn from that ensemble \textit{appears} to be a mix of devices and applications that a single host rarely runs, which could stand out from the norm.
Each user could instead keep a secret model trained on the user's own traffic, so that the user's \sysname flows resemble the user's own normal traffic.
A single user, however, produces far fewer flows than our training set holds, and a model of one user's traffic records that user's behavior, which makes the weights sensitive if they leak.
If the weights are public, on the other hand, every censor becomes a white-box censor (\S~\ref{sec:eval-quality}).
A possible middle ground, perhaps, is a public model that each deployment fine-tunes on its own, smaller sample of traffic and keeps secret.

\textbf{Drift and convergence.}
What's \textit{normal} changes over time and across contexts.
A model trained on one network at one time therefore becomes outdated, and a deployment needs to retrain it on recent traffic of its own network.
In the longer term, if \sysname's adoption reaches critical mass, \textit{will the flows of its users converge to a new shape?}
In that case, a deployment that retrains on recent traffic would then also train on the flows of other deployments (\ie on the output of earlier models).
Generative models trained on such output lose the tails of the original distribution over successive generations~\cite{shumailov2024collapse}.
Diversity across users---many models that each generate normal flows but differ from one another, as quality-diversity search seeks~\cite{pugh2016qd,mouret2015mapelites}---may counter such convergence.

\textbf{Conclusion.}
We have argued that the next generation of tunnel obfuscators should be schedule-driven, and that a good schedule blends into the traffic that's considered \textit{normal}.
\sysname realizes this design with a generative sequence model of such traffic.
We hope that \sysname and the problems that it leaves open draw more attention from the community and encourage more proposals and development around schedule-driven obfuscation.

\bibliographystyle{IEEEtran}
\bibliography{references}

\appendices
\section{Surveyed Obfuscators}
\label{app:survey}
Table~\ref{tab:survey} lists obfuscators for censorship circumvention that shape the sizes or timing of tunneled traffic.
We classified each obfuscator from its specification or source code in October 2026.
The list is not exhaustive, and it omits closed-source tools, which we could not inspect.
Tools that hide only the content of traffic, such as vanilla Shadowsocks, Trojan and VLESS, are out of scope.
A \textit{deployed} obfuscator has been shipped to users, and a \textit{prototype} has a public implementation without wide use.

Besides the two architectures of \S~\ref{sec:limits}, the table lists \textit{hybrid} obfuscators, which send some records without an application write but do not put all of their traffic on a schedule (\eg meek polls with no data while idle).

\begin{table}[t]
\centering
\caption{Obfuscators for censorship circumvention that shape the sizes or timing of tunneled traffic, by their architectural differences as described in \S~\ref{sec:limits}}
\label{tab:survey}
\footnotesize
\setlength{\tabcolsep}{4pt}
\begin{tabular}{@{}p{0.62in}p{0.72in}p{1.8in}@{}}
\toprule
Architecture & Status & Obfuscators \\
\midrule
Application-\allowbreak driven & Deployed (27) & \raggedright AnyTLS~\cite{anytlsgithub}, Cloak~\cite{cloakgithub}, fteproxy~\cite{dyer2013fte}, Geph~\cite{geph5github}, GoQuiet~\cite{goquietgithub}, Hysteria~2~\cite{hysteriagithub}, I2P~NTCP2/SSU2~\cite{i2pgithub}, MTProto~proxy~\cite{mtprotospec}, NaiveProxy~\cite{naiveproxygithub}, obfs2/3/4~\cite{lyrebirdgitlab,obfs4spec}, OQUIC~\cite{quicwrappergithub}, OSSH~\cite{psiphongithub}, Samizdat~\cite{samizdatgithub}, ScrambleSuit~\cite{winter2013scramblesuit}, Shadowsocks~2022~\cite{ss2022spec}, ShadowsocksR~plugins~\cite{ssrgithub}, Snell~\cite{singsnellgithub}, Sudoku~\cite{sudokugithub}, TapDance~\cite{wustrow2014tapdance}, TLSmasq~\cite{tlsmasqgithub}, Twiddle~\cite{twiddlegithub}, UDPspeeder~\cite{udpspeedergithub}, VMess~\cite{v2raygithub,xraygithub}, Windscribe~Stealth~\cite{windscribewstunnel}, XHTTP~\cite{xraygithub}, XTLS~Vision~\cite{xraygithub}, zapret2~wgobfs~\cite{zapret2github} \tabularnewline
 & Prototype (1) & \raggedright Scrambled~Shadowsocks~\cite{sssgithub} \tabularnewline
\midrule
Schedule-\allowbreak driven & Deployed (0) & \raggedright \textit{None} \tabularnewline
 & Prototype (8) & \raggedright BiFEP~\cite{chen2026bifep}, DeltaShaper~\cite{barradas2017deltashaper}, Marionette~\cite{dyer2015marionette}, Protozoa~\cite{barradas2020protozoa}, SkypeMorph~\cite{moghaddam2012skypemorph}, StegoTorus~\cite{weinberg2012stegotorus}, Stegozoa~\cite{figueira2022stegozoa}, Waterfall~\cite{nasr2017waterfall} \tabularnewline
\midrule
Hybrid & Deployed (3) & \raggedright meek~\cite{fifield2015fronting}, Restls~\cite{restlsgithub}, TLSMirror~\cite{v2raygithub} \tabularnewline
 & Prototype (1) & \raggedright basket2~\cite{yawning2016basket2} \tabularnewline
\bottomrule
\end{tabular}
\end{table}

\section{Server-Name Groups}
\label{app:sni}
\begin{figure}[t]
\centering
\includegraphics{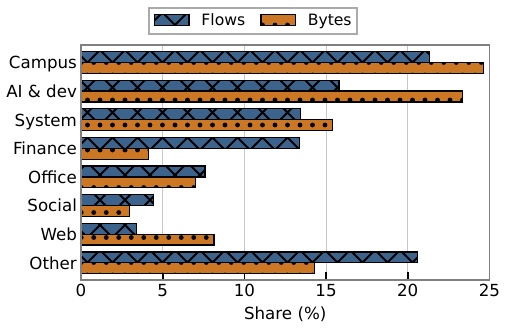}
\caption{Share of flows and bytes per server-name group.
AI \& dev: AI assistants, developer tools and cloud services.
System: OS updates, telemetry, analytics and network services.
Office: office suites, mail and storage.
Web: search, portals, media and shopping.
Other: names not identified and flows without a name.}
\label{fig:sni}
\end{figure}

In the TLS handshake, the client names the server that it wants to reach in the SNI extension; the flows of the dataset of \S~\ref{sec:dataset-data} name 29,017 distinct servers.
We assign each server name, by rules on the name and its registered domain, to one of 12 categories of service, or to Other if no rule applies, and merge the categories into seven groups (Fig.~\ref{fig:sni}).
Campus services, the campus's own servers, carry 21\,\% of the flows.
AI assistants, developer tools and cloud services carry 16\,\%.
System services and Finance carry 13\,\% each.
Office applications, mail and storage carry 8\,\%, messaging and social networks 4\,\%, and the Web group 3\,\%.
The remaining 21\,\% are flows whose server name no rule assigns (15\,\%) and flows without a server name (6\,\%).

\section{Tokenization}
\label{app:tokenizer}
This appendix details the quantizers $Q_s$ and $Q_t$ of \S~\ref{sec:model-repr}.
We fit both on the records of the training part that lie within the limits $L$ and $T$, and we discard a stream with fewer than three such records.
Algorithm~\ref{alg:tokenize} summarizes the tokenization of a flow.

\textbf{Size bins.}
$Q_s$ maps a signed size to a bin.
We fit it separately for the two directions, which share a budget of 200 bins in proportion to their records, with the boundary between them fixed at zero, so that each bin lies on one side of zero.
Within a direction, the fit has two stages.
First, every size that accounts for at least 5\,\% of the direction's records receives a bin of its own.
Second, the remaining sizes receive quantile bins, which hold about equal numbers of records and are therefore narrow where records are frequent.
The sizes of the first stage divide the direction into intervals, each interval receives a number of bins proportional to the number of records it holds, and the bin edges within an interval are its quantiles.
Because sizes are integers, a size that holds more records than one bin repeats as a quantile edge.
Such edges are merged, so the fit uses fewer bins than the budget: $B_s=153$ for the synchronous stream and $B_s=159$ for the asynchronous streams, whose sizes are owner-relative (\S~\ref{sec:model-dag}).

\textbf{Time bins.}
$Q_t$ maps a normalized IAT to a bin.
Quantile bins, which are narrow where values are frequent, do not suit IATs, because they would resolve differences below 1\,ms, where IATs reflect jitter and a sender in user space cannot schedule its writes reliably.
The first bin therefore holds all normalized IATs below 1\,ms, including zero.
Above 1\,ms, the bins have equal width in $\log\tilde\Delta$, four per decade, up to $T=30$\,s, which no tokenized IAT exceeds.
These logarithmic bins are too wide for IATs that concentrate near one value, as those of a timer do.
To find such peaks, we divide each logarithmic bin into 12 cells of about 0.02 decades (5\,\% wide) and count the IATs in each cell.
A cell belongs to a peak if it holds at least three times the share that it would hold if the IATs of its logarithmic bin were spread evenly, and at least 0.3\,\% of all IATs above 1\,ms; adjacent such cells form one peak.
Each peak at $c$ receives the bin $[c/1.05,\;1.05\,c]$, which replaces the part of the logarithmic bins that it covers and leaves room for timers that fire late within a tolerance (Apple recommends at least 10\,\% of the interval~\cite{apple-timer}).
On the training part, the fit finds peaks at 1.00\,s and 5.01\,s, and $Q_t$ has $B_t=22$ bins.

\textbf{Values within a bin.}
For every size bin, we store the sizes that it received in training with their relative frequencies, and for every time bin, a random sample of 1,000 of the normalized IATs that it received.
These are the empirical distributions $\pi_s(\cdot\mid s)$ and $\pi_t(\cdot\mid t)$ from which the generator draws.

\begin{algorithm}[t]
\caption{Tokenization of a flow (synchronous stream)}
\label{alg:tokenize}
\begin{algorithmic}[1]
\Require records $r_1,\dots,r_N$ in capture-point order; round trips $R_{\mathrm{s}},R_{\mathrm{c}}$; quantizers $Q_s,Q_t$; limits $L,T$
\Ensure token sequence $x$
\State $x\gets()$;\; $\Sigma\gets0$
\For{$i=1,\dots,\min(N,L)$}
  \If{$i=1$}
    \State $\tilde\Delta_i\gets0$
  \Else
    \State $\Delta_i\gets\tau_i^{\mathrm{s}}-\tau_{i-1}^{\mathrm{e}}$
    \State $\tilde\Delta_i\gets\max(0,\Delta_i-\hat\rho_i)$ \Comment{Eq.~\eqref{eq:norm}}
    \State $\Sigma\gets\Sigma+\tilde\Delta_i$
    \If{$\Sigma>T$} \textbf{break} \EndIf
  \EndIf
  \State append $\bigl(Q_s(d_ib_i),\,Q_t(\tilde\Delta_i)\bigr)$ to $x$
\EndFor
\If{the flow is complete}
  \State append $\mathrm{[EOF]}=(B_s,B_t)$ to $x$
\EndIf
\State \Return $x$
\end{algorithmic}
\end{algorithm}

\section{Sequence Model}
\label{app:model}
Here we detail the sequence model of \S~\ref{sec:model-arch}.
Both the synchronous and the asynchronous model consist of an input embedding, a backbone of Mamba blocks that summarizes the history $x_{\le i}$ in a hidden vector $\mathbf{h}_i\in\mathbb{R}^{d}$ with $d=256$, and two output heads (Fig.~\ref{fig:arch}).

\textbf{Input embedding.}
A token enters the model as its two bin indices, which are scaled to $[0,1]$ and mapped linearly to the model dimension:
\begin{equation}
\mathbf{u}_i=\mathbf{W}_{\mathrm{in}}\begin{bmatrix}s_i/B_s\\ t_i/B_t\end{bmatrix}+\mathbf{b}_{\mathrm{in}}+\mathbf{p}_i+\mathbf{r}_E,
\label{eq:app-embed}
\end{equation}
where $\mathbf{p}_i$ is the sinusoidal position code of the Transformer~\cite{vaswani2017attention}, and $\mathbf{r}_E$, which only the asynchronous model has, is a learned vector for the role of the endpoint $E$ that owns the stream (client or server).
Because the bins are ordered by size and by IAT, the scaled indices preserve this order, so records of similar size or IAT receive similar inputs.
We add the role vector at every position rather than supply it once as a prefix token, because in a recurrent model the influence of an early token can decay with distance, whereas an input added at every position reaches every step.

\textbf{Backbone.}
The backbone is a stack of Mamba blocks~\cite{gu2024mamba} followed by a final layer normalization that yields $\mathbf{h}_i$.
Mamba is a selective state-space model, which at each position $i$ updates a latent state $\mathbf{m}_i$ of fixed size from the preceding state and the current input $\mathbf{v}_i$, and computes its output $\mathbf{y}_i$ from the state:
\begin{equation}
\mathbf{m}_i=\bar{\mathbf{A}}_i\,\mathbf{m}_{i-1}+\bar{\mathbf{B}}_i\,\mathbf{v}_i,\qquad \mathbf{y}_i=\mathbf{C}_i\,\mathbf{m}_i .
\label{eq:app-ssm}
\end{equation}
In earlier state-space models, the matrices $\bar{\mathbf{A}}$, $\bar{\mathbf{B}}$ and $\mathbf{C}$ are fixed, so every input is processed alike.
Mamba instead computes them from the current input (\emph{selection}), so that at every step the model decides how much of its state to retain and how much of the input to write into it.
In a flow, the state can thus retain a request that a later response answers, while records that carry little information pass without displacing the request.
Each block normalizes its input, expands it to 512 channels, combines the four most recent positions with a short causal convolution, applies the recurrence of Eq.~\eqref{eq:app-ssm} with a state of 32 values per channel, scales the result by a gate computed from the input, projects it back to $d=256$ dimensions, and adds it to the input of the block.

Because the recurrence is linear in the state, training evaluates it at all positions of a sequence with a parallel scan, whose cost grows linearly with the sequence length, and generation evaluates it one step at a time on a state of fixed size.
The recurrence is also causal: the state at position $i$ depends only on the records up to $i$, so the model, like an endpoint, never represents a record with knowledge of the records that follow it.
A Transformer~\cite{vaswani2017attention}, in contrast, is not causal by construction: its attention relates every position to every other, and a causal mask must remove the later positions.
Even with the mask, a Transformer attends to all preceding tokens at every position, so its training cost grows quadratically with the sequence length, and in generation it keeps a cache that grows with every token, which makes the cost of each new record grow with the length of the flow.

\begin{figure*}[t]
\centering
\includegraphics{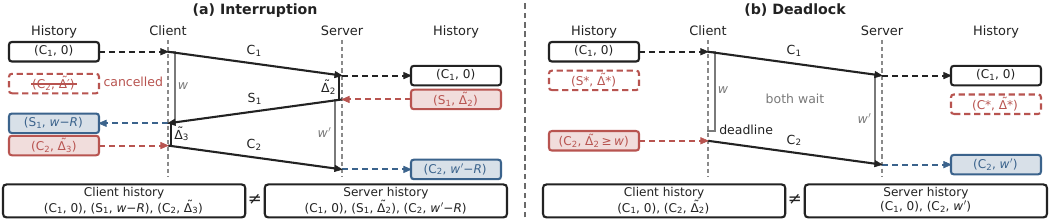}
\caption{The two race conditions of the asynchronous mode, with time running downwards. Red tokens are sampled by the model (dashed if not realized), blue tokens are computed from an arrival (Eq.~\eqref{eq:observe}), and C$_1$ (black) is the opener. $w$ and $w'$ are the times that the client and the server have waited since their last event, and Eq.~\eqref{eq:observe} subtracts $R$, the RTT estimated from the handshake, only if that event was the receiver's own send. (a)~The client samples a send C$_2$ with $\tilde\Delta'$, but S$_1$ arrives before $\tilde\Delta'$ has passed, so the client cancels the send, adds S$_1$ to its history and samples again. (b)~Both endpoints predict a receive and wait. When the client's deadline (Eq.~\eqref{eq:deadline}) passes, it decides again with $\tilde\Delta_2\ge w$ and sends C$_2$.}
\label{fig:async}
\end{figure*}

\textbf{Output heads.}
The model predicts the next token in two steps, its size bin first and then its time bin given the size bin (Eq.~\eqref{eq:factor}).
This factorization follows from the chain rule of probability and makes no independence assumption, since every joint distribution of a size bin and a time bin can be written in this form.
The size head is a linear layer followed by a softmax over the $B_s+1$ size classes, [EOF] included:
\begin{equation}
p_\theta(s_{i+1}=k\mid x_{\le i})=\operatorname{softmax}\bigl(\mathbf{W}_s\mathbf{h}_i+\mathbf{b}_s\bigr)_k .
\label{eq:app-sizehead}
\end{equation}
The time head receives, in addition to $\mathbf{h}_i$, a learned embedding $\mathbf{e}(s_{i+1})\in\mathbb{R}^{64}$ of the size bin and applies a multilayer perceptron with one hidden layer $\mathbf{q}_i\in\mathbb{R}^{d}$, followed by a softmax over the $B_t+1$ time classes:
\begin{equation}
\begin{aligned}
\mathbf{q}_i&=\operatorname{ReLU}\bigl(\mathbf{W}_1[\mathbf{h}_i;\mathbf{e}(s_{i+1})]+\mathbf{b}_1\bigr),\\
p_\theta(t_{i+1}=k\mid x_{\le i},s_{i+1})&=\operatorname{softmax}\bigl(\mathbf{W}_2\mathbf{q}_i+\mathbf{b}_2\bigr)_k .
\end{aligned}
\label{eq:app-timehead}
\end{equation}
Compared with a single softmax over joint tokens, the factorization has two advantages.
First, the joint vocabulary of the asynchronous model would have $B_sB_t+1=3{,}499$ classes, most of them rare, whereas its two heads have 160 and 23 classes and share their parameters across combinations.
Second, because the time head conditions on the size, the model represents how the IAT depends on the size of a record, and the hidden layer lets this dependence differ between histories, which a linear combination of the two inputs would not.
We predict the size first because the size bin determines the direction of the record, and with it the kind of transition that its IAT measures (\eg one that spans a round trip; \S\S~\ref{sec:model-repr} and~\ref{sec:model-dag}), and because [EOF], one of the size classes, decides whether a time is needed at all.

\textbf{Training.}
Training minimizes the mean NLL (Eq.~\eqref{eq:nll}) of the training tokens,
\begin{equation}
\mathcal{L}(\theta)=\frac{1}{|\mathcal{T}|}\sum_{(x,i)\in\mathcal{T}}\ell_i,
\label{eq:app-loss}
\end{equation}
where $\mathcal{T}$ contains every position $i\ge2$ of every training stream, sent and received records alike, so that the model also learns which endpoint sends next.
The time head is conditioned on the observed size bin (teacher forcing).
We use AdamW with a peak learning rate of $5\times10^{-4}$ and a weight decay of 0.01, a linear warm-up over 2,500 updates followed by cosine decay to a tenth of the peak, and gradient clipping at a norm of 1.0.
Because the lengths of the streams vary widely, a batch holds streams of similar length up to a total of 5,353 tokens rather than a fixed number of streams, so that little computation is spent on padding.
We divide the summed NLL of each batch by one constant rather than by its number of tokens, so that every token has the same weight.
We train the asynchronous model on the 46.4\,M asynchronous streams of the training part (507\,M tokens) and the synchronous model on its 23.2\,M synchronous streams (253\,M tokens), each for 20 epochs, and keep for each model the parameters with the lowest validation loss.

\textbf{Model scale.}
The depth of the backbone trades the accuracy of the model against its throughput.
To choose the depth, we trained asynchronous models of 6 to 21 Mamba blocks of the same width, each on training data in proportion to its depth (2.14\,M training streams per block, about 49 training tokens per parameter).
All depths used the same recipe, 10 epochs at a peak learning rate of $10^{-3}$, except that the 21-block model required $5\times10^{-4}$ to train stably.
We evaluated all models on the same 2.58\,M test streams (Fig.~\ref{fig:scale}).
The test NLL decreases from 1.310 at 6 blocks to 1.208 at 15 blocks and changes little beyond, reaching 1.201 at 21 blocks, while the scoring throughput on one GPU falls from 475\,k tokens per second at 6 blocks to 190\,k at 15 and 136\,k at 21.
We choose 15 blocks, which achieve 94\,\% of the improvement of 21 blocks over 6, the lowest NLL of all depths on short streams of at most 15 tokens (89\,\% of all streams), and 1.4 times the throughput of 21 blocks.
Beyond 15 blocks, the NLL improves only on long streams, which is consistent with the long-range dependence that motivated a sequence model (\S~\ref{sec:model-why}).
Both models used in the remainder of the paper have 15 blocks and 7.08\,M parameters (6.94\,M of them in the backbone), and are trained on the entire training part as described above.

\section{Perturbation Conditions and Concordance}
\label{app:validation}
This appendix details the validation of \S~\ref{sec:model-proxy}.

\textbf{Perturbation conditions.}
We draw 100,000 flows from the test part as the clean class.
\emph{Delete} removes a record and adds its IAT to that of the next record.
\emph{Insert} adds, after a record, a record in the same direction with a size drawn from the sizes of that direction and an IAT of 1\,ms.
\emph{Size jitter} multiplies the size of a record by $\exp(z)$ with $z\sim\mathcal{N}(0,0.2^2)$, and \emph{IAT jitter} multiplies its IAT by $\exp(z)$ with $z\sim\mathcal{N}(0,1.25^2)$.
We chose each standard deviation so that most jittered values leave their bin, which requires a larger value for the wider time bins: 84\,\% of the jittered sizes and 83\,\% of the jittered IATs of at least 1\,ms change their bin.
\emph{Swap} exchanges two consecutive records of the same sender, which leaves the set of tokens of each stream unchanged and alters only their order.
An action applies to each record independently with probability $p\in\{0.02,0.05,0.1,0.2,0.3,\dots,1\}$, identically in both streams of a flow.
We exclude deletion at $p=1$, which would remove every record, and obtain 59 conditions.

\textbf{Classifiers.}
Each classifier receives both asynchronous streams of a flow with their size and time tokens and is trained on up to 40,000 flows per class to separate the clean from the perturbed flows.
The distinguishability $D(\gamma)$ is the highest AUC that a classifier achieves on held-out flows.

\textbf{Distance metrics.}
For the NLL, the metric is the median flow NLL of the perturbed flows minus that of the same flows before perturbation.
Naive NLL$_2$ is the same difference under a bigram model of the tokens, $p(x_i\mid x_{i-1})$, fitted per stream role on the training part and interpolated with the token frequencies.
The other distances compare the token histograms of the clean and the perturbed flows: the total variation distance (TVD), the Jensen--Shannon divergence (JSD) and the Hellinger distance (HD) over single tokens and over pairs of consecutive tokens, and the Kullback--Leibler divergence (KL) over single tokens only.

\textbf{Concordance.}
A pair of conditions is comparable if its two conditions belong to different actions, both distinguishabilities lie in $[0.52,0.995]$, which excludes conditions at chance and saturated ones, and the distinguishabilities differ by more than twice the standard error of their difference.
Of the 59 conditions, 45 lie in this range, and they form 786 comparable pairs.
We exclude pairs within one action because $D$ grows with $p$ within an action, so any metric that also grows with $p$ agrees with $D$ there; across actions, a metric must place perturbations of different kinds on one common scale.
We obtain 95\,\% intervals by recomputing $C$ (Eq.~\eqref{eq:cindex}) for 500 draws of $D$ and of each metric from their estimated sampling distributions; every interval lies within $\pm0.05$ of its estimate, and the interval for the NLL is 0.951--0.958.

\section{Race Conditions of the Asynchronous Mode}
\label{app:races}
Fig.~\ref{fig:async} illustrates the two race conditions of the asynchronous mode (\S~\ref{sec:system-async}).

\section{Implementation}
\label{app:openscience}
\begin{figure}[t]
\centering
\resizebox{\columnwidth}{!}{\includegraphics{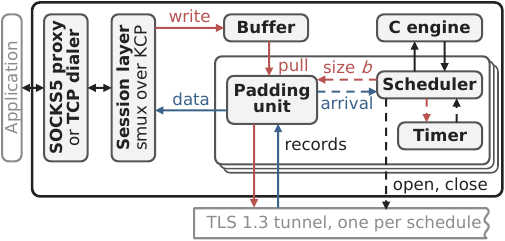}}
\caption{Components of one \sysname endpoint in the prototype. The stacked frame is repeated for every TLS tunnel; colours and line styles as in Fig.~\ref{fig:system}.}
\label{fig:impl}
\end{figure}

We implemented \sysname in Go, with the step of the model in the C engine of \S~\ref{sec:eval-res}, called through cgo; the source is available at \url{https://github.com/goose-lab-cuhk/obfsgen}.
Fig.~\ref{fig:impl} shows the components of one endpoint.
The client is a local SOCKS5 proxy, and the server dials the destination of each stream over TCP.

\textbf{Session layer.}
smux~\cite{smuxgithub} multiplexes the streams of the application over one KCP session~\cite{kcpgithub,kcpgogithub} per client, which provides reliable delivery.
KCP writes its packets into the buffer as if it were a datagram socket and receives the packets taken out of arriving records.
We disable its congestion control, since the schedule alone decides when data leaves, and enlarge its window to 4,096 packets, since acknowledgments return only in the records that the schedule of the peer calls for.

\textbf{TLS tunnels and records.}
Each TLS~1.3 tunnel plays one schedule and closes at [EOF].
A client keeps up to 16 tunnels open, which share its buffer, and the server groups the tunnels of a client by a session identifier.
TLS is configured so that every write produces exactly one record and that the server sends no record outside the schedule, such as a session ticket.
The plaintext of a record holds a flag byte, length-prefixed KCP segments and zero padding; a segment that does not fit is split, and its rest leaves in the next record of any tunnel.

\textbf{Opening a tunnel.}
The client takes $R$ from the TLS handshake, which spans one round trip between the endpoints, and draws a seed for the tunnel.
The size of the opener is drawn with this seed from the sizes of the first client records in the training part.
The opener carries a 36-byte hello with a session identifier, the seed, $R$ and an HMAC under a pre-shared key, by which the server authenticates the client, and a flag that selects the mode.
In the synchronous mode, both endpoints sample the plan from the seed, and the server closes the tunnel if the size of the opener does not match the seed; in the asynchronous mode, the server uses the $R$ of the client.

\textbf{Inference.}
Each tunnel keeps the recurrent state of its own stream, so that a decision costs one step of the engine, which all tunnels share and which runs int8 weights with AVX2, AVX-512 or NEON kernels, or in plain C.
The scheduler draws the random input of every decision from a seeded generator, so that both endpoints sample the same plan; the plans of 2,000 seeds were identical across the plain-C, AVX2 and AVX-512 builds.

\end{document}